\documentclass[useAMS,usenatbib]{mn2e}
\usepackage[fleqn]{amsmath}
\usepackage{graphicx}
\usepackage{threeparttable}
\usepackage{booktabs}
\usepackage{enumitem}
\usepackage{mathrsfs}
\title[Largest Bound Structures: The CSC]{The Largest Gravitationally Bound Structures: The Corona Borealis Supercluster - Mass and Bound Extent}
\author[D. W. Pearson, M. Batiste and D. J. Batuski]{David W. Pearson$^{1}$\thanks{E-mail: david.pearson@umit.maine.edu}, Merida Batiste$^{1}$\thanks{E-mail: merida.batiste@umit.maine.edu} and David J. Batuski$^{1}$ \\
$^{1}$ Department of Physics and Astronomy, University of Maine, 120 Bennett Hall, Orono, ME 04469, USA}
\begin{document}

\date{Accepted 2014 April 04. Received 2014 April 01; in original form 2014 February 17}

\pagerange{\pageref{firstpage}--\pageref{lastpage}} \pubyear{2013}

\maketitle

\label{firstpage}

\begin{abstract}
Recent simulations of the densest portion of the Corona Borealis supercluster (A2061, A2065, A2067, and A2089) have shown virtually no possibility of extended gravitationally bound structure without inter-cluster matter (Pearson \& Batuski). In contrast, recent analyses of the dynamics found that the clusters had significant peculiar velocities towards the supercluster centroid (Batiste \& Batuski). In this paper we present the results of a thorough investigation of the CSC: we determine redshifts and virial masses for all 8 clusters associated with the CSC; repeat the analysis of Batiste \& Batuski with the inclusion of A2056 and CL1529+29; estimate the mass of the supercluster by applying the virial theorem on the supercluster scale (e.g. Small et al.), the caustics method (e.g. Reisenegger et al.), and a new procedure using the spherical collapse model (SCM) with the results of the dynamical analysis (SCM+FP); and perform a series of simulations to assess the likelihood of the CSC being a gravitationally bound supercluster. We find that the mass of the CSC is between \mbox{$0.6$ and $12 \times 10^{16} \, h^{-1} \, \mathrm{M}_{\sun}$}. The dynamical analysis, caustics method and the SCM+FP indicate that the structure is collapsing, with the latter two both indicating a turn around radius of \mbox{$\sim 12.5 \, h^{-1} \, \mathrm{Mpc}$}. Lastly, the simulations show that with a reasonable amount of inter-cluster mass, there is likely extended bound structure in the CSC. Our results suggest that A2056, A2061, A2065, A2067, and A2089 form a gravitationally bound supercluster.
\end{abstract}


\section{Introduction}
Superclusters of galaxies are the largest coherent structures in the Universe, and studies \citep{Rood76} have shown that their internal dynamics are generally dominated by Hubble flow (i.e. not gravitationally bound). Extended bound structure within superclusters is unusual, but intensive studies of the Shapley Supercluster (SSC) (e.g. \cite{Bardelli93,Proust06}, and references therein) have demonstrated significant bound structure, with a central core of five clusters that is in the final stages of collapse \citep{Reisenegger00,Munoz08,Pearson13}. To date the SSC is the only confirmed bound supercluster in the Universe (for the purposes of the current paper we define a supercluster as containing no fewer than $3$ rich clusters of galaxies), but its existence suggests that similar structures might be found elsewhere in the Universe.

The Corona Borealis supercluster (CSC) is a particularly dense, compact supercluster that has been identified as a candidate bound supercluster similar to the SSC. It was included by \citet{Abell61} in his catalog of second order clusters and first noted by \citet{Shane59}, who initially identified 12 member clusters in a \mbox{$6\degr \times 6\degr$} region before later using brightest cluster galaxies to show that there were actually two components viewed in projection. The foreground component, at \mbox{$z\approx 0.07$}, contains Abell clusters 2056, 2061, 2065, 2067, 2079, 2089 and 2092 and is what we now refer to as the CSC. \citet{Postman88} performed the first dynamical analysis of this region, making virial mass estimates of six clusters (excluding A2056), finding a mass for the supercluster of \mbox{$8.2 \times 10^{15} \, h^{-1} \, \mathrm{M}_{\sun}$}. The Norris Survey \citep{small1,small2,small3} expanded on this work, performing N-body simulations to test mass estimators, demonstrating that the virial theorem was applicable on supercluster scales. By treating the six clusters in the core as a single virialized system, they estimated a mass of at least \mbox{$3 \times 10^{16} \, h^{-1} \, \mathrm{M}_{\sun}$}, concluding that the system was bound and had likely reached turnaround. \citet{Kopylova98} performed a dynamical analysis of the CSC using redshift independent distance estimates to assess peculiar velocities for eight clusters in the region (including A2019 and A2124 in addition to those of Postman et al.). Their results led them to define a rapidly collapsing core containing five clusters; A2061, A2065, A2067, A2089 and A2092.

Recently \cite{Pearson13} (hereafter P13) and \cite{Batiste13} (hereafter B13) have revisited these results with the intention of more accurately assessing the current dynamical state of the CSC using currently available observational data. P13 performed N-body simulations of the CSC using the most accurate cluster mass estimates available for A2061, A2065, A2067 and A2089 (and assuming negligible inter-cluster mass), concluding that there is very little likelihood that any part of the structure is bound. B13 used Sloan Digital Sky Survey (SDSS) data \citep{SDSS7} to perform a dynamical analysis with the six clusters used by Postman et al., finding peculiar velocities indicative of extended bound structure. 

In this paper we aim to address the apparent conflict between these results, providing a more complete picture of the current dynamical state of this structure, and a reasonable estimate of the mass of the bound portion. We undertake an extensive analysis, employing several independent methods of mass estimation, and performing a large number of simulations with which we can assess the validity of the conclusions we draw from the dynamical analysis. The breadth of the analysis is intended to provide a context within which the affects of the assumptions underlying each method can be assessed, and the significant uncertainties on each result can be interpreted.

The paper is structured as follows: In section 2 we describe the different data sets used for mass estimation and dynamical analysis, and explain the selection criteria and corrections to photometry. In Section 3 we present the dynamical analysis. In section 4 we present the mass estimation methods, which constrain the mass for several different scenarios: upper and lower bounds are placed on the possible mass of the CSC; the mass required to generate the motions reported in Section 3 is determined; and an alternative assessment of the mass is made using an independent set of observational data. In section 5 we present the results of our simulations. We use these results to assess the likelihood that the CSC is a bound structure, and to investigate whether inter-cluster dark matter would need to contribute significantly to the total supercluster mass. Lastly, in section 6 we discuss the implications of our results.

Throughout this paper we adopt a standard \mbox{$\Lambda\mathrm{CDM}$} cosmology with \mbox{$H_{0}=100 \, h \, \mathrm{km} \; \mathrm{s}^{-1} \, \mathrm{Mpc}^{-1}$}, \mbox{$h = 0.7$}, \mbox{$\Omega_{\Lambda,0} = 0.7$}, \mbox{$\Omega_{m,0} = 0.3$} and \mbox{$q_{0}=-0.55$}.  

\section{Observational Data}
\label{TheCSC}
\subsection{Data for positions and mass estimation}
\label{sec:data:mass}
\begin{figure*}
\includegraphics[width=0.49\textwidth]{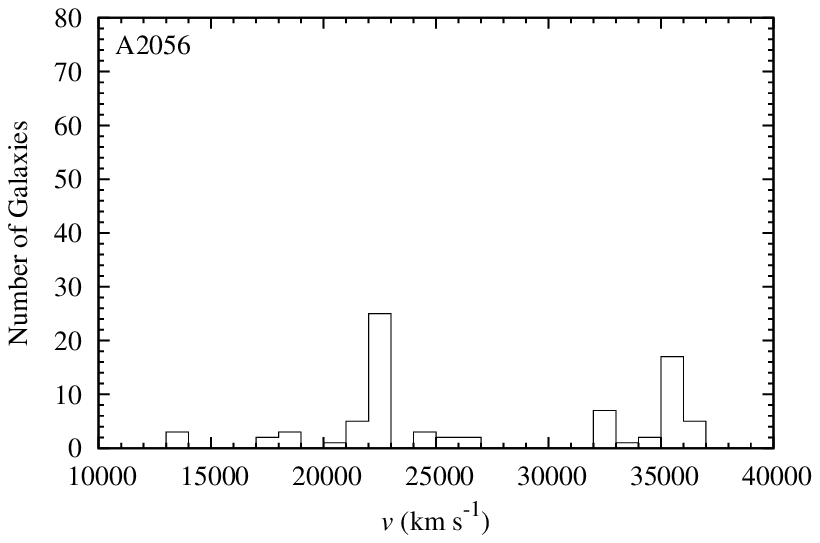} \hfill
\includegraphics[width=0.49\textwidth]{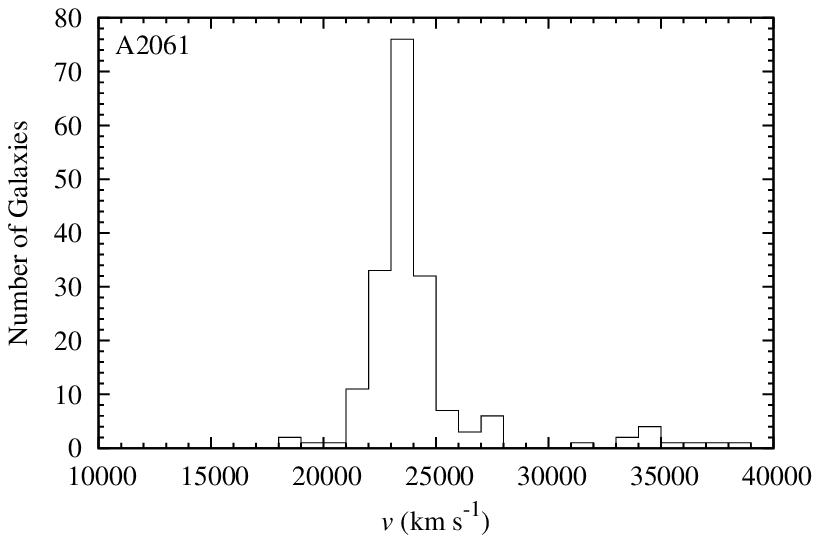}
\includegraphics[width=0.49\textwidth]{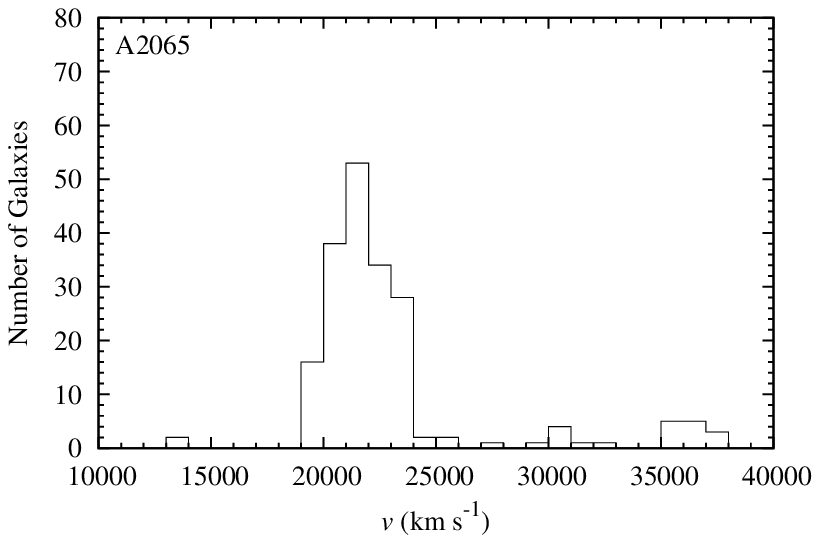} \hfill
\includegraphics[width=0.49\textwidth]{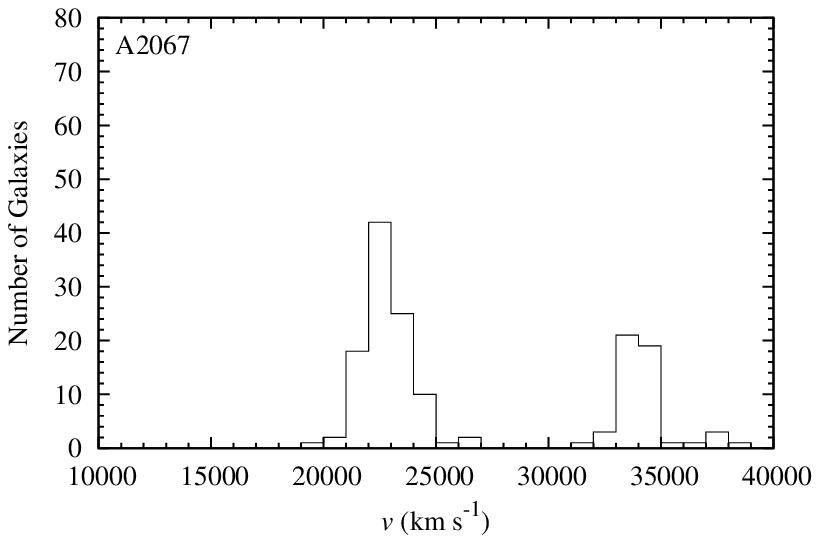}
\includegraphics[width=0.49\textwidth]{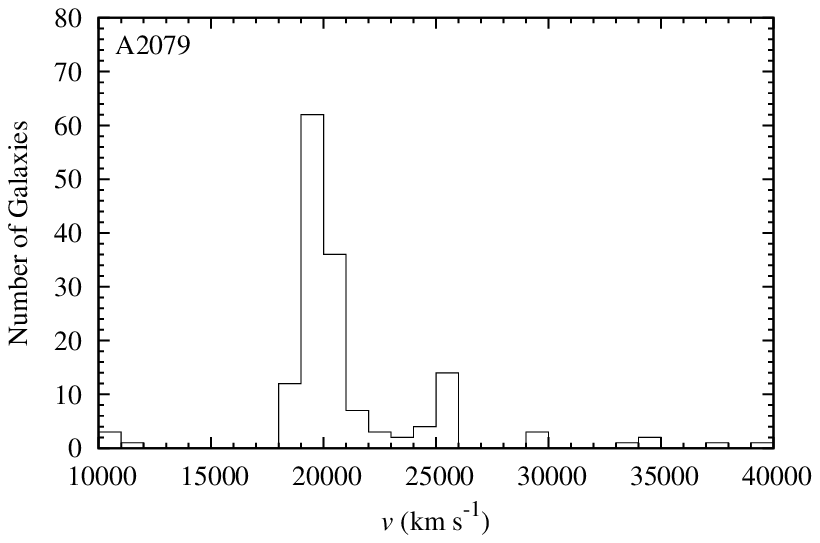} \hfill
\includegraphics[width=0.49\textwidth]{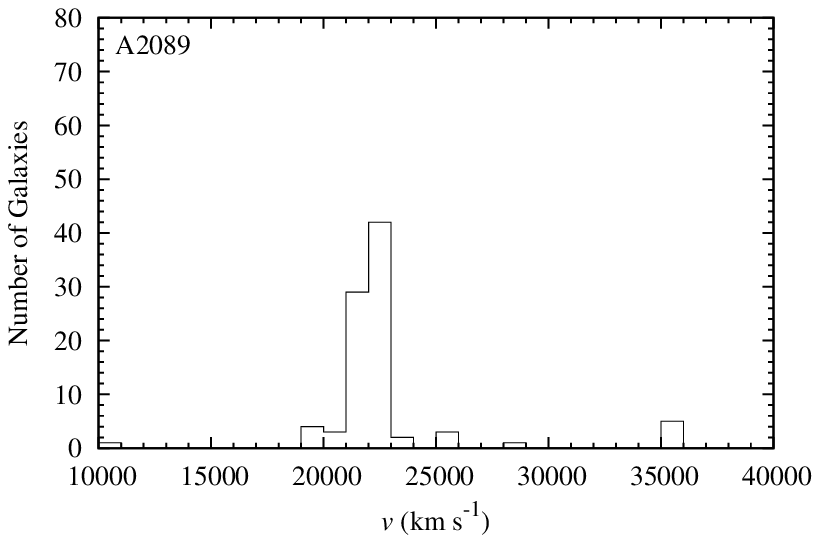}
\includegraphics[width=0.49\textwidth]{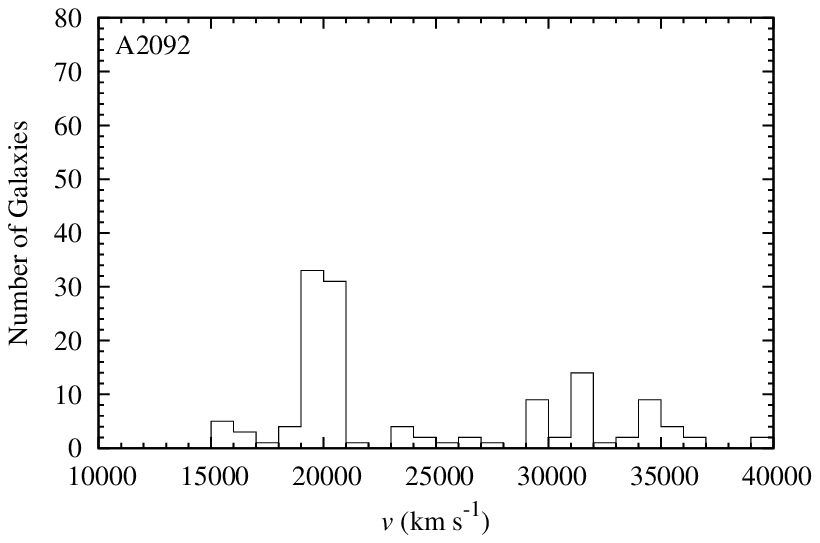} \hfill
\includegraphics[width=0.49\textwidth]{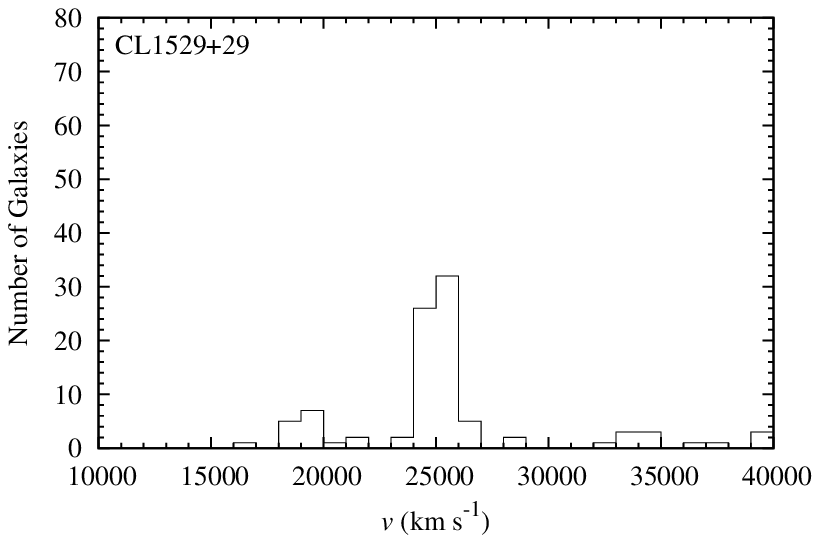}
\caption{Histograms of the redshifts within $1.5 \, h^{-1} \, \mathrm{Mpc}$ of the cluster centres for the members of the CSC.}
\label{ClusterHist}
\end{figure*}

With the goal of identifying gravitationally bound structure in the CSC, we choose to focus on the particularly dense region containing the clusters A2056, A2061, A2065, A2067, A2079, A2089, A2092, and CL1529+29. We begin by searching the NASA/IPAC Extragalatic Database (NED) for all galaxies with measured redshifts in the range \mbox{$0.03 < z < 0.13$} in a \mbox{$\sim 6\degr \times 6\degr$} region centered on \mbox{$15^{\mathrm{h}}24^{\mathrm{m}}$} right ascension and \mbox{$+30\degr$} declination in J2000 coordinates. With the recent completion of two large redshift surveys, the SDSS and the 2 Micron All Sky Survey (2MASS) Redshift Survey \citep{Huchra12}, NED contains large quantities of galaxies with measured redshifts in the region. Starting from the cluster coordinates and redshifts listed in NED, we locate all galaxies within \mbox{$1.5 \, h^{-1} \, \mathrm{Mpc}$} of the cluster centres and, based on the histograms (see Fig. \ref{ClusterHist}), set initial redshift limits for each cluster. We then apply an iterative $3\sigma$ clipping procedure to remove interlopers. With the remaining galaxies we estimate the cluster centroid positions, and redshifts using the biweight method of \cite{Beers90}. We also calculate the velocity dispersions \citep{Carlberg96}
\begin{equation}
\label{disp}
\sigma^{2} = \left(\sum_{i} w_{i}\right)^{-1} \sum_{i} w_{i} \left(\Delta v_{i}\right)^{2},
\end{equation}
where $w_{i}$ is the weighting factor, and \mbox{$\Delta v_{i}$} is the difference between a galaxy's redshift and the centroid value. We estimate the harmonic radius from the ringwise estimator \citep{Carlberg96}
\begin{equation}
\label{r_h}
r_{h}^{-1} = \left(\sum_{i} w_{i}\right)^{-2} \sum_{i<j} \dfrac{w_{i}w_{j}}{2\mathrm{\pi}} \int_{0}^{2\mathrm{\pi}} \dfrac{d\theta}{\sqrt{r_{i}^{2} + r_{j}^{2} + 2r_{i}r_{j}\cos(\theta)}},
\end{equation}
where $r_{i}$ and $r_{j}$ are the projected radii of the $i$th and $j$th galaxies. Combining that information, we estimate the masses with the virial theorem
\begin{equation}
\label{MVir}
M_{\mathrm{V}} = \dfrac{3\mathrm{\pi}\sigma^{2}r_{h}}{2G}.
\end{equation}

Table \ref{CMSep} lists the results of the above analysis. Error estimates in the velocities, dispersions, and masses are from bootstrap resampling. The error estimates immediately next to the values are the 68 per cent, and those in parentheses are the 95 per cent confidence intervals. Both are listed since the upper and lower limits at the 68 per cent level for each cluster are averaged together and used as $1\sigma$ to generate a Gaussian distribution, from which we can randomly draw mass realizations for our simulations.

\begin{figure}
\begin{center}
\includegraphics{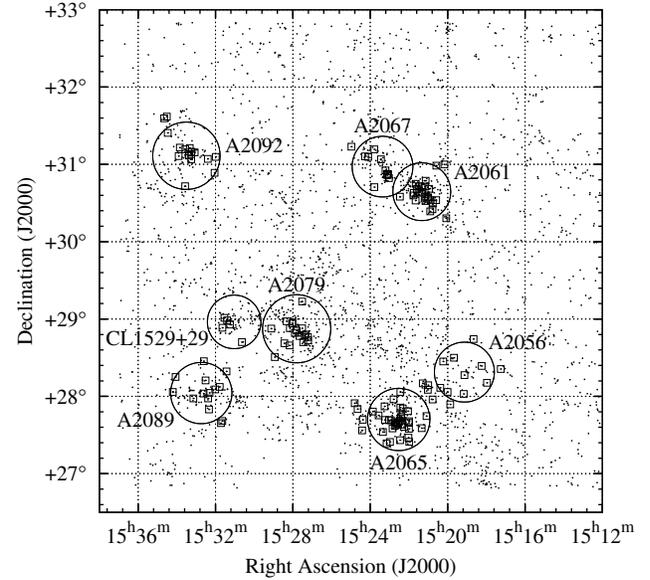}
\caption{Positions on the sky of the CSC clusters, with all galaxies used for mass estimation (small dots) and the FP analysis (open squares) shown. Coordinates for the cluster centres are given in Table \ref{tab:FPclust}. The circles have radius equivalent to $1.5\:h^{-1}$ Mpc at the redshift of the cluster.}
\label{fig:clusterpos}
\end{center}
\end{figure}

In Fig. \ref{fig:clusterpos}, we plot the locations of all the clusters from Table \ref{CMSep} with circles representing a \mbox{$1.5 \, h^{-1} \, \mathrm{Mpc}$} radius at the cluster redshift. The small black dots show the locations of a subset of galaxies \mbox{($0.052 < z < 0.098$)} with measured redshifts that we collected from NED, showing that the region is fairly well covered.

\begin{table*}
\centering
\caption{The clusters of the CSC. Column (1) lists the cluster number. Columns (2) and (3) list the right ascension and declination of the cluster centers in J2000 coordinates, respectively. Column (4) gives the number of galaxies that were used in the calculation. Columns 5 and 6 list the recessional velocities and dispersions, respectively. Column (7) lists the harmonic radius of the cluster. Lastly, column (8) lists the mass estimate.}
\label{CMSep}
\begin{tabular}{lcccccc}
\hline
Cluster & RA & Dec. & $N$ & $v$ & $\sigma$ & $M$ \\
${}$ & (J2000) & (J2000) & ${}$ & ($\mathrm{km} \, \mathrm{s}^{-1}$) & ($\mathrm{km} \, \mathrm{s}^{-1}$) & ($10^{15} \, h^{-1} \, \mathrm{M}_{\sun}$) \\
\hline
A2056 & 15$^{\mathrm{h}}$19$^{\mathrm{m}}$08.1$^{\mathrm{s}}$ & $+28\degr 18\arcmin 50\arcsec$ & 30 & $22,525_{-107}^{+58} \left(_{-189}^{+142}\right)$ & $350_{-63}^{+32} \left(_{-111}^{+80}\right)$ & $0.186_{-0.084}^{+0.021} \left(_{-0.137}^{+0.074}\right)$ \\ [1.2ex]
A2061 & 15$^{\mathrm{h}}$21$^{\mathrm{m}}$17.6$^{\mathrm{s}}$ & $+30\degr 38\arcmin 33\arcsec$ & 152 & $23,508_{-78}^{+73} \left(_{-154}^{+148}\right)$ & $898_{-59}^{+51} \left(_{-114}^{+106}\right)$ & $0.990_{-0.153}^{+0.099} \left(_{-0.277}^{+0.216}\right)$ \\ [1.2ex]
A2065 & 15$^{\mathrm{h}}$22$^{\mathrm{m}}$31.8$^{\mathrm{s}}$ & $+27\degr 42\arcmin 04\arcsec$  & 172 & $21,673_{-104}^{+94} \left(_{-193}^{+203}\right)$ & $1286_{-63}^{+56} \left(_{-123}^{+115}\right)$ & $2.330_{-0.318}^{+0.163} \left(_{-0.559}^{+0.405}\right)$ \\ [1.2ex]
A2067 & 15$^{\mathrm{h}}$23$^{\mathrm{m}}$21.9$^{\mathrm{s}}$ & $+30\degr 58\arcmin 18\arcsec$ & 73 & $22,247_{-112}^{+94} \left(_{-215}^{+197}\right)$ & $771_{-81}^{+70} \left(_{-157}^{+145}\right)$ & $0.885_{-0.217}^{+0.125} \left(_{-0.388}^{+0.296}\right)$ \\ [1.2ex]
A2079 & 15$^{\mathrm{h}}$27$^{\mathrm{m}}$48.3$^{\mathrm{s}}$ & $+28\degr 52\arcmin 30\arcsec$ & 117 & $19,767_{-76}^{+76} \left(_{-152}^{+152}\right)$ & $720_{-61}^{+44} \left(_{-114}^{+97}\right)$ & $0.805_{-0.157}^{+0.080} \left(_{-0.275}^{+0.198}\right)$ \\ [1.2ex]
A2089 & 15$^{\mathrm{h}}$32$^{\mathrm{m}}$43.9$^{\mathrm{s}}$ & $+28\degr 02\arcmin 32\arcsec$ & 75 & $22,082_{-64}^{+63} \left(_{-128}^{+126}\right)$ & $519_{-39}^{+33} \left(_{-75}^{+69}\right)$ & $0.414_{-0.082}^{+0.038} \left(_{-0.142}^{+0.098}\right)$ \\ [1.2ex]
A2092 & 15$^{\mathrm{h}}$33$^{\mathrm{m}}$30.2$^{\mathrm{s}}$ & $+31\degr 06\arcmin 43\arcsec$ & 69 & $19,988_{-64}^{+51} \left(_{-122}^{+109}\right)$ & $526_{-52}^{+45} \left(_{-100}^{+93}\right)$ & $0.308_{-0.072}^{+0.041} \left(_{-0.128}^{+0.098}\right)$ \\ [1.2ex]
CL1529+29 & 15$^{\mathrm{h}}$31$^{\mathrm{m}}$02.1$^{\mathrm{s}}$ & $+28\degr 57\arcmin 53\arcsec$ & 64 & $25,182_{-106}^{+69} \left(_{-194}^{+156}\right)$ & $586_{-56}^{+41} \left(_{-104}^{+89}\right)$ & $0.468_{-0.130}^{+0.062} \left(_{-0.225}^{+0.157}\right)$ \\
\hline
\end{tabular}
\end{table*}

The exact redshift of A2056 from the literature is questionable, with early measurements placing it at 0.0763 \citep{Struble87,ACO} based on the measurement of a single galaxy reported in \cite{Fairall71}. Later, \cite{Struble99} reported a redshift of 0.0834, again based on a single measured galaxy. The Norris survey only obtained 10 redshifts for A2056, so the authors decided to exclude it from their analysis. Looking at the top left panel of Fig. \ref{ClusterHist}, we see the reason for contradictory redshifts of A2056 in the literature. There is a strong peak in galaxy count at \mbox{$22,500 \, \mathrm{km} \; \mathrm{s}^{-1}$} \mbox{($z \sim 0.075$)}, and a smaller collection of galaxies immediately behind the peak around \mbox{$25,000 \, \mathrm{km} \; \mathrm{s}^{-1}$} \mbox{($z \sim 0.083$)}. The value we obtain, \mbox{$z = 0.0751$} \mbox{($v = 22,525 \, \mathrm{km} \; \mathrm{s}^{-1}$)}, places A2056 in the core of the CSC very near A2065, the most massive cluster in the region. Given that the redshift is based on 30 measured galaxies, we feel confident that it is part of the structure, but we also note that the sampling of the cluster is fairly uniform and does not contain many measurements near the core (see Fig. \ref{fig:clusterpos}), likely due to a lack of pointed observations. Looking at Fig. \ref{ClusterHist}, it can be seen that there is a significant concentration of galaxies in a background supercluster around \mbox{$35,000 \, \mathrm{km} \; \mathrm{s}^{-1}$}, along the same line-of-sight as A2056. This, along with the relatively low mass estimate listed in Table \ref{CMSep}, may imply that A2056 is only a richness class 0 cluster.

\subsection{Data for the dynamical analysis}
\label{sec:obs:data}
The dynamical analysis of B13 used the Fundamental Plane relation (FP) for early-type galaxies (ETGs) \citep{Djorgovski87} to determine accurate redshift independent distances to six clusters: A2061, A2065, A2067, A2079, A2089, and A2092. In this work we extend that analysis to include the additional clusters A2056 and CL1529+29. Our calculated spectroscopic redshift for CL1529+29 (Figure \ref{ClusterHist}) suggests that it is unlikely to be part of the bound structure, but we include it for completeness and also as a test of the robustness of our method. 

Data for each cluster are taken from the Seventh Data Release of the SDSS. Details of the selection criteria and corrections to photometry are given in B13, so we describe them only briefly here. 

Initial cluster positions and redshifts are taken from NED and each cluster is given a radius of 30 arcmin on the sky and an initial redshift range of $\pm0.015$ about the centre. Iterative $3\sigma$ clipping is applied to identify cluster members, and the final quoted cluster redshifts (Table \ref{tab:FPclust}) come from a biweight average of the cluster member redshifts. 

Selection criteria, based on those of \cite{Bernardi03a}, are applied to each of the cluster samples to select ETGs for the FP analysis:

\begin{enumerate}[leftmargin=0.5cm,label={(\arabic*)},itemsep=3pt]
\item
\mbox{$FRACDEV\geq 0.8$}
\item
\mbox{$\frac{b}{a} \geq 0.6$}
\item
\mbox{$eclass < 0$}
\item
\mbox{$\sigma \geq 100\, \mathrm{km} \; \mathrm{s}^{-1}$}
\item
\mbox{$14.50 \leq r^{*} \leq 17.45$ }
\end{enumerate}

The first three criteria are intended to reject galaxies with a disc component: (1) requires that the galaxy light profile be best fitted by a de Vaucouleurs $r^{1/4}$ model, (2) applies an axis ratio cut \citep{BinneyMerri}, and (3) uses a PCA classification to reject galaxies with emission line spectra, characteristic of late-type disc galaxies. The spectral resolution of the SDSS means that velocity dispersion estimates lower than \mbox{$90 \, \mathrm{km} \; \mathrm{s}^{-1}$} are considered unreliable, so (4) applies a slightly more conservative cut at \mbox{$100 \, \mathrm{km} \; \mathrm{s}^{-1}$}. This has the additional effect of reducing the scatter in our best fitting FP, as ETGs with velocity dispersions lower than \mbox{$100 \, \mathrm{km} \; \mathrm{s}^{-1}$} have been shown to systematically deviate from the FP that best fits those with higher dispersions \citep{Gargiulo09}. Finally, the magnitude limits (5) are based on the completeness of the sample, and match those of \cite{Bernardi03a}. Table \ref{tab:FPclust} gives the spectroscopic redshifts, and sample sizes for each cluster, after these selection criteria have been applied.

Corrections to photometry are made as described in B13, and include the corrections to magnitudes and effective radii of \citet{Hyde09}, which account for the poor sky subtraction in the SDSS data reduction pipeline. The mean effective surface brightness is corrected for cosmological dimming and K-corrected following \cite{kcor}
\begin{equation}
\langle\mu\rangle_{e} = m+2.5\log\left(2 \pi r_{e}^{2}\right)-K(z)-10\log(1+z).
\end{equation} 

\begin{table*}
\caption{Details of the FP data sets. Column (2) shows spectroscopic redshifts, calculated from a biweight average of the galaxy redshifts. Columns (3) and (4) give the number of galaxies used in the cluster redshift determinations and in the FP analysis. Columns (5) and (6) give the best fitting FP coefficients for each cluster. All errors are calculated via a boot-strapping procedure. Values for the supercluster are shown in the last line.}
\label{tab:FPclust}
\begin{tabular}{lccccc}
\hline 
Cluster & $z_{spec}$ & $N_{spec}$ & $N_{FP}$ & $a$ & $b$ \\ 
\hline 
A2056 & $0.0758\pm 0.0007$ & 56 & 15 & $0.75\pm 0.27$ & $0.290\pm 0.042$ \\
A2061 & $0.0788\pm 0.0002$ & 83 & 25 & $1.50\pm 0.33$ & $0.351\pm 0.034$ \\
A2065 & $0.0717\pm 0.0004$ & 152 & 38 & $1.10\pm 0.12$ & $0.359\pm 0.023$ \\ 
A2067 & $0.0740\pm 0.0002$ & 50 & 11 & $1.18\pm 0.78$ & $0.270\pm 0.190$ \\ 
A2079 & $0.0659\pm 0.0003$ & 83 & 18 & $0.99\pm 0.21$ & $0.321\pm 0.061$ \\ 
A2089 & $0.0737\pm 0.0003$ & 64 & 15 & $1.14\pm 0.37$ & $0.360\pm 0.054$ \\ 
A2092 & $0.0664\pm 0.0002$ & 59 & 16 & $1.10\pm 0.26$ & $0.332\pm 0.034$ \\ 
CL1529+29 & $0.0856\pm 0.0007$ & 8 & 5 & -- & -- \\ [2ex]
CSC & $0.0725\pm 0.003$ & -- & 142 & $1.08\pm 0.07$ & $0.327\pm 0.019$ \\
\hline
\end{tabular}
\end{table*}

Velocity dispersions are aperture corrected following \cite{Jorgensen95}. In what follows, $r_{e}$ represents effective radius in angular units and $R_{e}$ is in physical units of $\mathrm{h}^{-1}\:\mathrm{kpc}$, and all quantities have all been corrected following the methods described above.

\section{Dynamical Analysis}
\label{sec:dynamics}
\subsection{The Fundamental Plane as a Distance Indicator}
\label{sec:dynamics:FP}
The FP relates three structural and kinematic properties of ETGs, and is given by
\begin{equation} \label{eq:FP}
\log r_{e}= a\log\sigma + b\langle\mu\rangle_{e} + c,
\end{equation} 
where $\sigma$ is the line-of-sight velocity dispersion, $r_{e}$ is effective radius, which contains half the total light of a galaxy, and $\langle\mu\rangle_{e}$ is mean effective surface brightness, which is given by
\begin{equation}
\langle\mu\rangle_{e} \propto -2.5\log\left(\frac{L}{2\pi r_{e}^{2}}\right),
\end{equation}
where $L$ is the luminosity. When effective radius is measured in angular units, the zero point offset $c$ is directly dependent on the distance of a galaxy, and this relation can be used to make redshift-independent distance determinations. The FP is analogous to the Tully-Fisher relation for disc galaxies \citep{TF} and has similar dispersion in the fit, equivalent to \mbox{$\sim20$ per cent} error in distance to a single galaxy. Averaging offsets for a number of galaxies in a single cluster allows this same method to be used to determine cluster distances with significantly better accuracy, since errors in cluster distance estimates scale as $N^{1/2}$ \citep{Gibbons01}.

If all clusters are fit to the same FP, then equation \eqref{eq:FP} can be shown to give:
\begin{equation} \label{eq:distance}
\dfrac{D_{2}}{D_{1}}=10^{c_{1}-c_{2}}
\end{equation}
where $D_{1}$ and $c_{1}$ are the distance and associated offset for a chosen calibrator, and $D_{2}$ and $c_{2}$ are the distance and associated offset for the cluster of interest. Thus, calibration is essential in determining accurate absolute cluster distances, and the accuracy of the FP distance determinations must be viewed as being relative to that of the calibration. 

\subsection{Fitting the FP}
\label{sec:dynamics:fitting}
The coefficients that best describe the FP vary widely in the literature, so we find the coefficients $a$ and $b$ that best fit the data for all clusters in our sample. There are a number of methods by which it is possible to determine a best fitting FP, but we choose to perform a direct fit in \mbox{$\log r_{e}$}, minimizing residuals in the distance dependent parameter. In order to account for possible bias, which may result from the intrinsic dispersion in the FP as well as measurement error and correlated errors among the observables, we use the Bivariate Correlated Errors and Intrinsic Scatter fit \citep{Akritas96}. This method employs an average covariance matrix to correct the results of a least-squares fit, providing an unbiased estimate of the best fitting coefficients (complete details of the fitting procedure can be found in B13). This method will not account for any bias due to selection effects, but this should not be a concern in using the FP as a distance indicator, since the whole sample will be affected by the selection criteria in the same way. Columns (7) and (8) of Table \ref{tab:FPclust} give the best fitting $a$ and $b$ coefficients for each of the clusters, as well as the global best fitting coefficients for the supercluster, with associated errors. 

\begin{figure}
\includegraphics[width=1.0\linewidth]{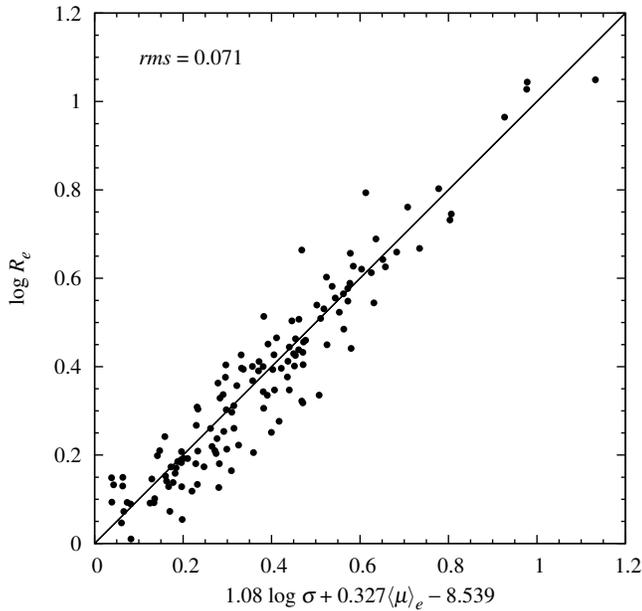}
\caption{Best fitting FP for all 142 galaxies in the sample, with $R_{e}$ in \mbox{$h^{-1}\:\mathrm{kpc}$}.}
\label{fig:FP}
\end{figure}

It should be noted that the amount of data available for CL1529+29 is smaller than for any other cluster in the sample. This stems at least partly from the magnitude limits that we have used: once the other selection criteria have been applied, this magnitude limit leaves only 5 ETGs. Given the small sample size it is not possible to make a good determination of the best fitting FP for this cluster, so we do not include those coefficients either in Table \ref{tab:FPclust} or in our determination of the global best fitting plane. Instead, we fit those data to the plane that best describes the other clusters, which is given by

\begin{equation} \label{eq:FP_best_fit2}
\log R_{e} = 1.08\log\sigma+0.327\langle\mu\rangle_{e}-8.539.
\end{equation}
B13 found a best fitting plane of: 
\begin{equation} \label{eq:FP_best_fit}
\log R_{e} = 1.11\log\sigma+0.339\langle\mu\rangle_{e}-8.853.
\end{equation}
Comparing equations \eqref{eq:FP_best_fit2} and \eqref{eq:FP_best_fit} we see that the inclusion of A2056 does not significantly affect the fit, with the values for $a$ and $b$ in both cases being consistent within the errors. The best fitting FP for all galaxies in the sample, corrected to the same inertial reference frame, is shown in Figure \ref{fig:FP}. The dispersion in the fit is $0.071$, which is effectively identical to the dispersion of $0.069$ in the fit of B13, and is equivalent to \mbox{$\sim16$ per cent} error in distance to a single galaxy.  

\subsection{Distances and Peculiar Velocities}
\label{sec:dynamics:Dist}
Since the FP does not provide a direct measure of distance, it is necessary to calibrate the relation. The magnitude limits on the SDSS preclude using a nearby cluster with a known distance, such as the Coma cluster, for this calibration. Instead we assume that the clusters are, on average, at rest with respect to the CMB (i.e. no peculiar motion for the supercluster centroid). We then use the biweight method to iteratively reduce both the cluster redshifts and the cluster offsets to the supercluster centroid, thereby providing values of $D_{1}$ and $c_{1}$ that can be used with equation \eqref{eq:distance}. The offset for the supercluster ($c_{1}$) is given in equation \eqref{eq:FP_best_fit2}, and the redshift of the supercluster centroid (from which we obtain $D_{1}$) is shown in Table \ref{tab:FPclust}. 

Cluster distances, determined from equation \eqref{eq:distance}, are converted to $z_{FP}$ using \citep{Peebles93}
\begin{equation}
D=\dfrac{c z}{H_{0}} \left(1- \dfrac{(1+q_{0})z}{2}\right) 
\end{equation}
and are then compared with the spectroscopic cluster redshifts to assess the peculiar velocity for each cluster \citep{Danese80}
\begin{equation}\label{eq:V_pec}
V_{pec}=c \left( \dfrac{ z_{spec}-z_{FP}}{1+z_{FP} }\right).  
\end{equation}

Table \ref{tab:FP} shows the offsets $c$ for each cluster, the FP distances ($z_{FP}$), and the resulting peculiar velocities. Given that CL1529+29 is unlikely to be part of the bound structure, and that the calibration offset does not change significantly when this cluster is included, we choose to exclude CL1529+29 from our calculation of the calibration redshift.

\begin{table*}

\caption{Results from the FP analysis, using the global best fitting FP given by equation \ref{eq:FP_best_fit2}. Column (2) gives the cluster offsets, calculated from a biweight average of the galaxy offsets for each cluster, with errors calculated via a boot-strapping procedure. Column (3) gives the rms dispersion in the fits for each cluster. Column (4) gives the cluster distances determined from the FP. Column (5) gives the error in the distance estimates, determined from the rms dispersion in the global fit. Column (6) gives the resulting peculiar velocities.}
\label{tab:FP}
\begin{tabular}{lccccc}
\hline 
Cluster & $c$ & $rms$ & $z_{FP}$ & error & $V_{pec}$ \\
	&	&	&	& ($\%$)& $(\mathrm{km}\:\mathrm{s}^{-1})$ \\
\hline 
A2056 & $-8.529\pm0.017$ & $0.078$ & 0.0720 & 4 & 1074 \\
A2061 & $-8.533\pm0.017$ & $0.070$ & 0.0727 & 3 & 1708 \\
A2065 & $-8.535\pm0.010$ & $0.065$ & 0.0730 & 3 & -355 \\
A2067 & $-8.531\pm0.031$ & $0.070$ & 0.0723 & 5 &  476 \\ 
A2079 & $-8.483\pm0.019$ & $0.078$ & 0.0646 & 4 &  366 \\  
A2089 & $-8.530\pm0.026$ & $0.069$ & 0.0722 & 4 &  415 \\ 
A2092 & $-8.516\pm0.018$ & $0.063$ & 0.0698 & 4 & -963 \\ 
CL1529+29 & $-8.585\pm 0.062$ & -- & 0.0820 & 7 & 993 \\
\hline 
\end{tabular}
\end{table*}

The largest peculiar velocity is indicated for A2061, and there is a significant relative peculiar velocity between A2061 and A2067 \mbox{($\sim 1200 \, \mathrm{km} \; \mathrm{s}^{-1}$)}. Given their proximity along the line-of-sight, as indicated by the FP distances, this large relative peculiar velocity is likely due to these clusters being a closely bound pair. \cite{Marini04} used X-ray data to investigate the possibility that this was a merging pair and found evidence that A2061 is significantly elongated towards A2067. The FP distances (column (4) of Table \ref{tab:FP}) suggest that this elongation might reasonably be due to a gravitational interaction between the clusters. The significant difference between their peculiar velocities should also be mentioned, as we cannot exclude the possibility that this difference is an artefact of incomplete sampling. Given the overlap between the clusters on the sky and their proximity along the line-of-sight, we applied quite a narrow redshift range to each cluster to avoid contamination between the samples (see discussion in B13). As a result, the samples are certainly not complete in volume and the elongation of A2061 is not well represented. B13 used several methods to test the robustness of the distance estimates to variations in the samples, as well as applying a different secondary distance indicator to analyze the dynamics. While the resulting individual peculiar velocities for A2061/67 varied somewhat, a significant relative peculiar velocity of \mbox{$1200-1400\, \mathrm{km} \; \mathrm{s}^{-1}$} was consistently present between them. We conclude from this that, while the calculated peculiar velocities for each cluster may be affected by our sampling, the relative motion is accurately represented.

A2065 is the most massive of the clusters and its small peculiar velocity is consistent with it being the central mass concentration for the structure. The peculiar velocity of A2089, as well as its position relative to A2065, suggests they are likely bound. The results for A2092 suggest it may also be part of bound structure, but has not yet reached turnaround. Given its position on the sky relative to the more massive clusters, if it is part of bound structure, then it is likely that it is bound to the pair A2061/67. 

The FP distance of A2079 suggests that it is likely foreground and not part of the bound structure. However, for reasons detailed in B13 and outlined in Section \ref{sec:dynamics:systematics}, we treat this result with caution and suggest that it cannot be entirely excluded from future analysis of this structure. 

The FP distance of A2056 indicates that it is very close to A2065 along the line-of-sight. Given their proximity on the sky and the substantial mass of A2065, the significant peculiar velocity of A2056 is consistent with these clusters being bound.
  
The FP distance of CL1529+29 places it well beyond the CSC core, and significantly more distant than any of the other clusters, supporting the conclusion that it is not gravitationally bound to the CSC core clusters.

The distances and peculiar velocities determined for the six clusters of B13 do not change meaningfully with the inclusion of A2056 or CL1529+29. The possible exception to this is A2079, for which the largest difference can be seen in the FP distance determination compared with B13. The inclusion of A2056 in the analysis results in a lower peculiar velocity for A2079, which is consistent with no significant peculiar motion relative to the supercluster centroid. 

\subsection{Systematic Errors and Biases}
\label{sec:dynamics:systematics}
B13 investigated possible sources of systematic error and bias among the cluster samples and the effects that those might have on the results from the FP. They found that the only significant potential sources of error arose from the sparse and variable sampling of the clusters, and some intrinsic differences between the cluster populations that may well be a consequence of the sampling differences. We perform a similar analysis here, including A2056 but not CL1529+29. Given the small sample size and the higher mean effective surface brightness, we do expect to see some systematic differences for CL1529+29. However, since the inclusion of this cluster does not affect the distance estimates for the other clusters, and since we are confident that it is not part of the bound region, our overall conclusions are not affected by these differences.

The cluster samples are all drawn from the same source and treated consistently, so no bias arises from differences in methods of data acquisition or reduction. Great care has been taken to avoid bias in the best fitting FP due to measurement error and contamination from disc or low-velocity dispersion galaxies. We do not account for bias due to selection effects, as these will affect the whole sample similarly, and since we are using the FP as a relative distance indicator instead of trying to measure the ``true" FP, this should not introduce any bias in the distance estimates.

Following the procedure of B13 we test the assumption, inherent in our use of the FP, that all ETGs are drawn from the same underlying distribution. Anderson-Darling (AD) k-sample tests are performed with all the galaxy properties used in the FP (\mbox{$\log R_{e}$}, \mbox{$\log \sigma$}, \mbox{$\langle\mu\rangle_{e}$}, and the offsets $c$) as well as subsets of those properties. The AD test is chosen instead of the more commonly used Kolmogorov-Smirnoff (KS) test because it is more sensitive to variations between distributions \citep{Hou09}. We perform a seven sample test in each observable, finding that the null hypothesis of a single underlying distribution is accepted in all cases except for \mbox{$\langle\mu\rangle_{e}$}, where it is rejected with \mbox{$P=0.015$}. Considering the joint distribution in \mbox{$\log R_{e}$}, \mbox{$\log \sigma$} and \mbox{$\langle\mu\rangle_{e}$} we find that the null hypothesis is rejected with \mbox{$P=0.04$}. These results are similar to those of B13 and indicate some intrinsic differences among the clusters, so we perform two-sample tests among all the possible pairs of clusters. As in B13, we find that A2079 is involved in almost all cases where the null hypothesis is rejected, and when it is excluded from the k-sample test the null hypothesis is accepted in both \mbox{$\langle\mu\rangle_{e}$} \mbox{($P=0.52$)} and the joint distribution of \mbox{$\log R_{e}$}, \mbox{$\log \sigma$} and \mbox{$\langle\mu\rangle_{e}$} \mbox{($P=0.54$)}. 

These results indicate that the A2079 sample systematically deviates from the others, particularly regarding the higher mean central surface brightness. B13 showed that, though this systematic bias is clearly present, it does not serve to bias the distance estimates of the other clusters in the analysis. This suggests that, while the results for A2079 should be viewed with caution, the statistical analysis does not indicate any bias in the overall results as a consequence of the assumption that all ETGs are drawn from the same underlying distribution.

\section{Supercluster Mass Estimation}
\label{scmasssec}
The oft cited \cite{small3} suggests that there is evidence of a very large amount of mass in the CSC region. By applying the virial theorem on the supercluster scale, which they justify from the results of N-body simulations, they determine that the mass present must be somewhere between \mbox{$3.3$ and $\sim 8\times 10^{16} \, h^{-1} \mathrm{M}_{\sun}$}. This result should be viewed with some caution, since applying the virial theorem essentially assumes that the structure is bound, and thus will simply give the amount of mass required to bind the supercluster. In order to test the robustness of their results, and inform our simulations that include an inter-cluster matter component, we apply several different methods of estimating supercluster masses to the CSC: We repeat the calculations of \cite{small3} using a much larger and more uniformly sampled data set; we apply the caustics method used by \cite{Reisenegger00} on the Shapley supercluster; and lastly, we use the spherical collapse model (SCM) in concert with the FP results (section \ref{sec:dynamics:Dist}) to obtain a mass estimate which is, to our knowledge, the first application of such a procedure to any set of objects. By applying a number of different methods we can establish upper and lower bounds on the mass, as well as perform independent checks of each method. We discuss these procedures in detail in the following sections.

\subsection{Application of the virial theorem}
\label{SCVir}
\begin{figure}
\centering
\includegraphics[width=1.0\linewidth]{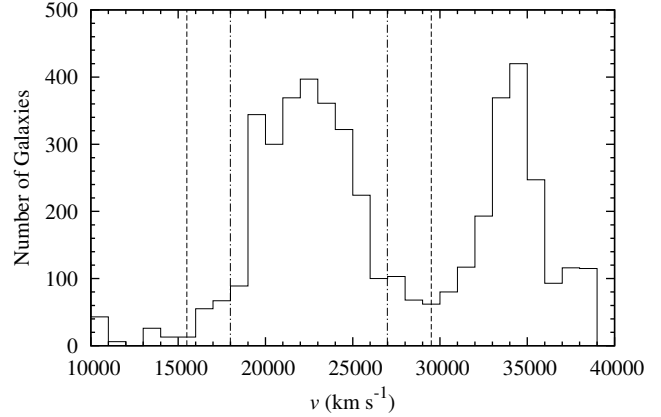}
\caption{A histogram of all redshifts in the CSC region available in NED from \mbox{$0.03 < z < 0.13$}, with bins of \mbox{$1000 \, \mathrm{km} \; \mathrm{s}^{-1}$}. The dot-dashed vertical lines are the limits adopted by Small et al. 1998, while the dashed vertical lines are the less stringent limits adopted here to set a firm upper limit on the total mass of the supercluster with the virial theorem.}
\label{CSCHist}
\end{figure}

The data from NED (see section \ref{sec:data:mass}) gives us \mbox{$\sim 10,000$} galaxies with measured redshifts. Before applying the virial theorem, we need to establish the boundaries of the supercluster in redshift space. For this purpose, we plot a histogram of all redshifts obtained from NED in Fig. \ref{CSCHist}. We note that there are two peaks in the histogram, one centered at \mbox{$\sim 22,500 \, \mathrm{km} \; \mathrm{s}^{-1}$} (the CSC) and one at \mbox{$\sim 35,000 \, \mathrm{km} \; \mathrm{s}^{-1}$} (the background supercluster). Given the shape of the CSC peak we choose to apply two different redshift space limits to obtain mass estimates using equations \eqref{disp}, \eqref{r_h} and \eqref{MVir}.

The first limits applied are identical to those used in \cite{small3} \mbox{($0.06 < z < 0.09$)} and are represented by the dot-dashed vertical lines in Fig. \ref{CSCHist}. Within these limits we have a total of 2,503 galaxies, giving a bi-weight median recessional velocity of \mbox{$22,357_{-102}^{+92} \, \mathrm{km} \; \mathrm{s}^{-1}$}, and a dispersion of \mbox{$2,121_{-43}^{+43} \, \mathrm{km} \; \mathrm{s}^{-1}$}. These lead to a mass estimate  of \mbox{$6.63_{-0.31}^{+0.26} \times 10^{16} \, h^{-1} \, \mathrm{M}_{\sun}$} for the CSC. The uncertainties on these numbers are the 95 per cent confidence limits from bootstrap resampling.

The second set of limits applied are the widest possible \mbox{($0.052 < z < 0.098$)} that we could reasonably fit to the structure, giving us a firm upper bound on the mass. These limits are shown in Fig. \ref{CSCHist} as dashed vertical lines, and within them we find 2,836 galaxies. This gives us a new biweight centroid velocity of \mbox{$22,474_{-129}^{+124} \, \mathrm{km} \; \mathrm{s}^{-1}$}, and a new dispersion of \mbox{$2,760_{-75}^{+65} \, \mathrm{km} \; \mathrm{s}^{-1}$}, leading to a mass of \mbox{$11.6_{-0.7}^{+0.6} \times 10^{16} \, h^{-1} \, \mathrm{M}_{\sun}$}.

As in \cite{small3}, we also estimate the mass based solely on the disperson amongst the clusters. This reduces our sample to eight, leading to a significantly higher uncertainty for this result. We arrive at a biweight centroid velocity of \mbox{$22,068_{-882}^{+1052} \, \mathrm{km} \; \mathrm{s}^{-1}$}, with a dispersion of \mbox{$1,642_{-690}^{+507} \, \mathrm{km} \; \mathrm{s}^{-1}$}, leading to a mass \mbox{$3.7_{-2.9}^{+1.4} \times 10^{16} \, h^{-1} \mathrm{M}_{\sun}$}. 

The results using only the clusters, and limits identical to \cite{small3} are entirely consistent with their results. Based on the results of N-body simulations, they conclude that, given their sparse sampling of the region, they may underestimate the mass by a factor of 2 leading them to set an upper limit of \mbox{$\sim 8 \times 10^{16} \, h^{-1} \, \mathrm{M}_{\sun}$}. Looking at Fig. \ref{fig:clusterpos} we can see that our data set does not suffer from the same sparse sampling, meaning that the mass estimate we obtain should be more accurate. That result \mbox{($6.63_{-0.31}^{+0.26} \times 10^{16} \, h^{-1} \, \mathrm{M}_{\sun}$)} is within the estimated upper bound of \cite{small3}. For the cluster-only analysis they arrived at \mbox{$\sim 4 \times 10^{16} \, h^{-1} \, \mathrm{M}_{\sun}$} using 7 clusters (all the ones listed in Table \ref{CMSep} minus A2056).

\subsection{Application of the SCM+FP}
\begin{figure*}
\includegraphics[width=0.48\linewidth]{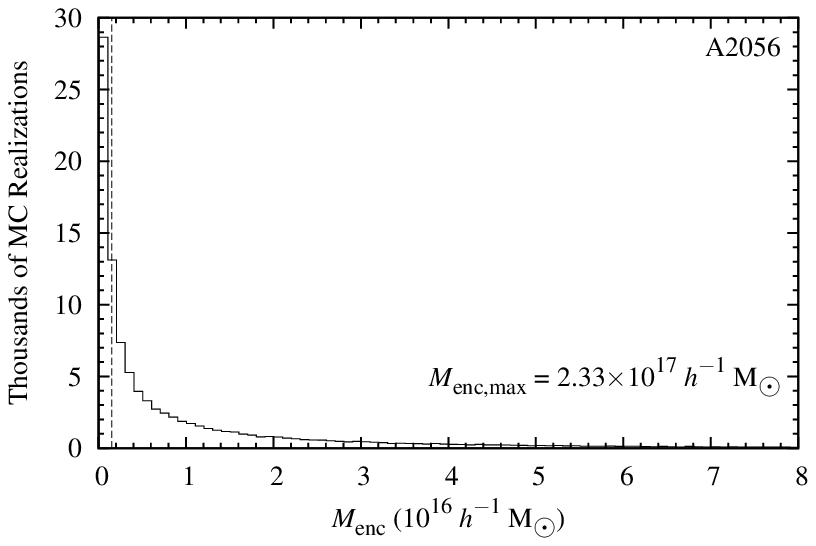}
\includegraphics[width=0.48\linewidth]{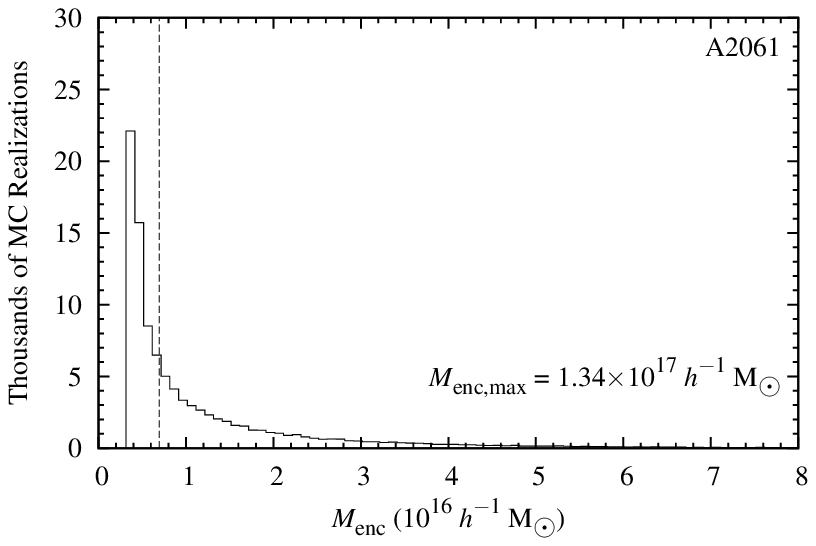}
\includegraphics[width=0.48\linewidth]{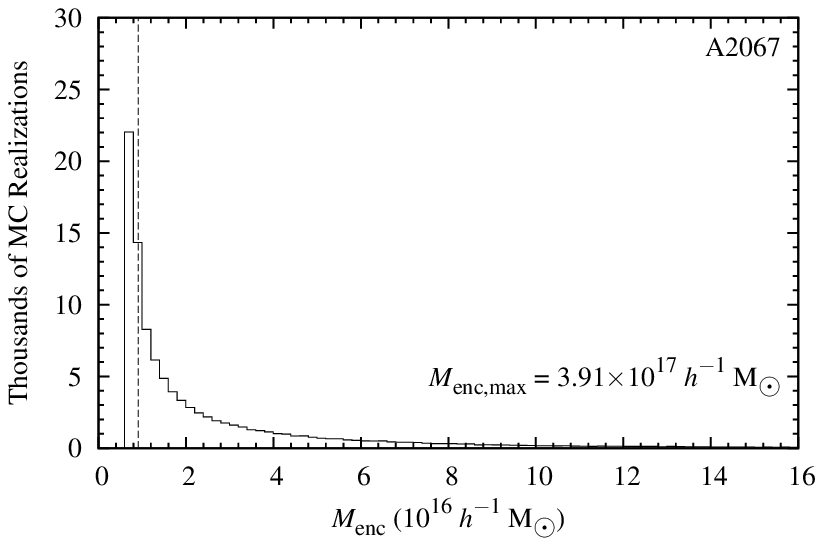}
\includegraphics[width=0.48\linewidth]{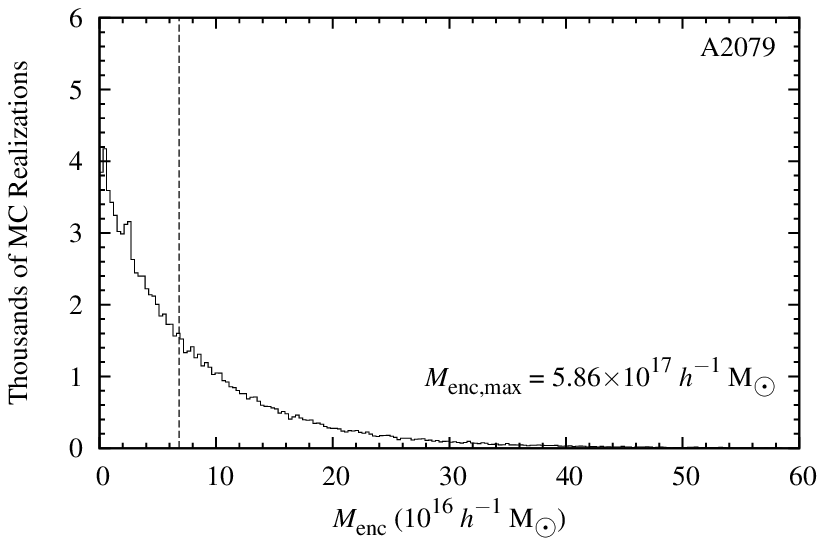}
\includegraphics[width=0.48\linewidth]{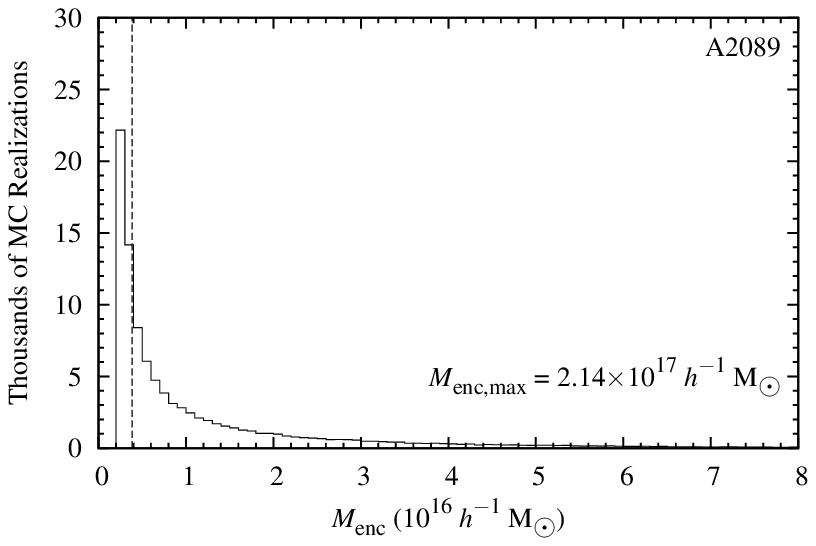}
\includegraphics[width=0.48\linewidth]{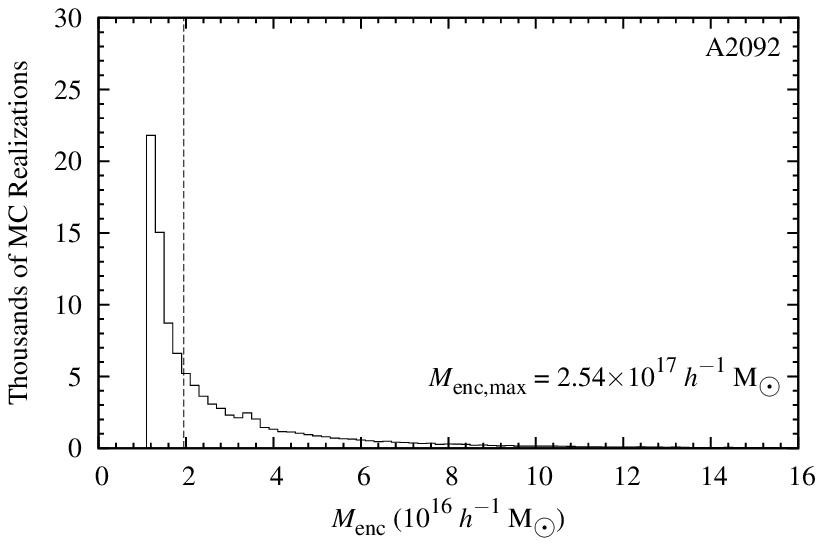}
\includegraphics[width=0.48\linewidth]{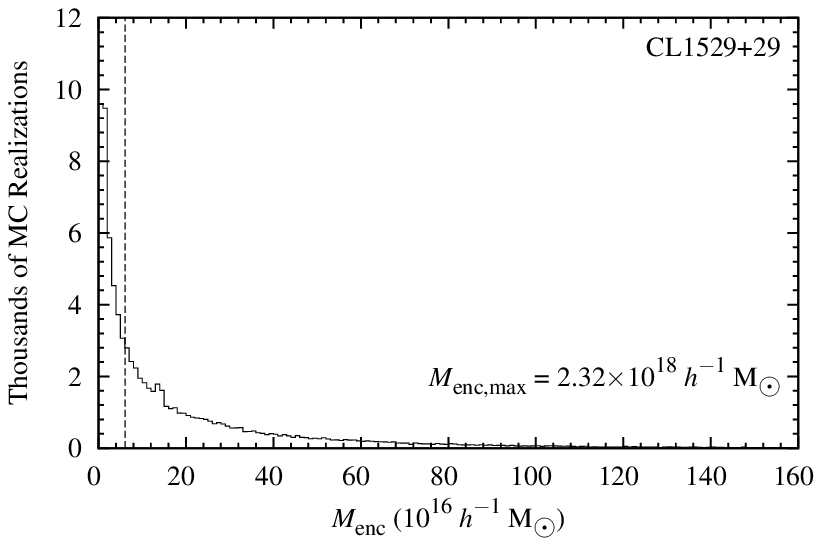}
\caption{Histograms of the masses resulting from the MC realizations of the $z_{FP}$ for each cluster. The high-mass tails of the distribution extend quite far, so in an effort to make the graphs more readable, the range has been truncated and the maximum value from the MC realizations is listed on the plot for each cluster.}
\label{MCHist}
\end{figure*}

\begin{table*}
\centering
\caption{Results of applying the SCM+FP method to the CSC. Column (1) lists the cluster name. Column (2) is the separations of the clusters from A2065. Column (3) lists the radial velocity relative to A2065. Column (4) lists the velocity parameter, with column 5 listing the density parameter. Column (6) gives the density inside the shell, and column (7) lists the enclosed mass implied by the density parameter.}
\label{SCM+FPTab}
\begin{tabular}{lcccccc}
\hline
Cluster & $\Delta r_{\mathrm{A2065}}$ & $v_{r}$ & $A$ & $\Omega_{\mathrm{s}}$ & $\rho_{\mathrm{enc}}$ & $M_{\mathrm{enc}}$ \\
${}$ & ($h^{-1} \, \mathrm{Mpc}$) & ($\mathrm{km} \; \mathrm{s}^{-1}$) & ${}$ & ${}$ & ($10^{12} \, h^{2} \, \mathrm{M}_{\sun} \, \mathrm{Mpc}^{-3}$) & ($10^{16} \, h^{-1} \, \mathrm{M}_{\sun}$) \\
\hline
A2056 &	4.65 &	-666 &	2.0488 &	13.08 &	3.63 &	0.153 \\
A2089 &	8.97 &	-186 &	0.0429 &	4.59 &	1.27 &	0.385 \\
A2061 &	11.23 &	-154 &	0.0188 &	4.22 &	1.17 &	0.695 \\
A2067 &	12.51 &	-118 &	0.0089 &	4.00 &	1.11 &	0.910 \\
A2092 &	18.00 &	306 &	0.0288 &	2.88 &	0.80 &	1.952 \\
A2079 &	23.68 &	-419 &	0.0313 &	4.43 &	1.23 &	6.835 \\
CL1529+29 &	27.54 &	710 &	0.0664 &	2.51 &	0.70 &	6.094 \\
\hline
\end{tabular}
\end{table*}

Given accurate redshift independent distances, we can perform a more robust estimation of the mass needed to generate the observed motions of the clusters. In order to do this, we begin with the spherical collapse model, since it is well studied and known to give remarkably accurate results for its simplicity. The energy associated with a shell in this model in a $\Lambda\mathrm{CDM}$ universe is given by
\begin{equation}
\label{SCE}
E = \dfrac{1}{2}\left(\dfrac{dr}{dt}\right)^{2} - \dfrac{GM_{\mathrm{enc}}}{r} - \dfrac{1}{6}\Lambda r^{2},
\end{equation}
where $r$ is the radius of the shell, and $M_{\mathrm{enc}}$ is the mass enclosed. It is convenient to transform this equation into dimensionless quantities defined by \citep{D06}
\begin{equation}
\label{dlr}
\widetilde{r} = \left(\dfrac{\Lambda}{3GM_{\mathrm{enc}}}\right)^{1/3} r,
\end{equation}
\begin{equation}
\label{dlt}
\widetilde{t} = \left(\dfrac{\Lambda}{3}\right)^{1/2} t,
\end{equation}
\begin{equation}
\label{dlE}
\widetilde{E} = E\left(\dfrac{(GM_{\mathrm{enc}})^{2}\Lambda}{3}\right)^{-1/3}.
\end{equation}
The dimensionless energy is then
\begin{equation}
\label{twidleE}
\widetilde{E} = \dfrac{1}{2}\left(\dfrac{d\widetilde{r}}{d\widetilde{t}}\right)^{2} - \dfrac{1}{\widetilde{r}} - \dfrac{\widetilde{r}^{2}}{2}.
\end{equation}

From equation \eqref{twidleE}, we can setup the integral equation
\begin{equation}
\label{tint}
\int_{0}^{t_{0}} dt = \int_{0}^{\widetilde{r}_{0}} \dfrac{\sqrt{\widetilde{r}} \, d\widetilde{r}}{\sqrt{\widetilde{r}^{3} + 2\widetilde{E}\widetilde{r} +2}},
\end{equation}
which can only be solved analytically for the special case of the critical shell energy, \mbox{$\widetilde{E} = -3/2$} \citep{D06}, but it can of course be evaluated numerically. Since we are working in dimensionless units, we perform a single set of numerical integration which can then be applied to any structure with the transformations defined in equations \eqref{dlr}--\eqref{dlE}. The left hand side simply gives the time you wish to integrate to, while the results from the right hand side are the radius of the shell with a given energy at that time. To integrate equation \eqref{tint}, we select energies starting at \mbox{$\widetilde{E} = 0$}, decreasing iteratively, and carrying out the integration until the result of the right hand side gives the current dimensionless time. We keep going until we find the shell with \mbox{$\widetilde{r} = 0$} (\mbox{$\widetilde{E} = -2.2$}).  We note that the results of our numerical integration agree well with those of \cite{D06} yielding a value of \mbox{$\widetilde{r}_{\mathrm{TA}} = 0.73$} for the turn-around energy of \mbox{$\widetilde{E}_{\mathrm{TA}} = -1.64$}, as well as a value of \mbox{$\widetilde{r}_{\mathrm{cs}} = 0.84$} for the critical energy of \mbox{$\widetilde{E}_{\mathrm{cs}} = -1.5$}.

\cite{D06} show that the density parameter of a shell is given by
\begin{equation}
\label{DenParamS}
\Omega_{\mathrm{s}} = \dfrac{2\Omega_{\Lambda}}{\widetilde{r}^{3}},
\end{equation}
and since the numerical integration gives us the radii of the shells at present, we calculate their associated density parameters. The density parameters are tied to the peculiar velocities indicated by the FP analysis through the velocity parameter,
\begin{equation}
\label{VelParam}
A = \left(\dfrac{1}{H_{0}r}\dfrac{dr}{dt}\right)^{2}.
\end{equation}
\cite{D06} showed that in terms of the energy and density parameters, the velocity parameter can be calculated as
\begin{equation}
\label{ACalc}
A = \Omega_{\Lambda} + \Omega_{\mathrm{s}} + \widetilde{E}(\Omega_{\mathrm{s}};\Omega_{\Lambda})\left(2\Omega_{\Lambda} \Omega_{s}^{2}\right)^{1/3}.
\end{equation}
Thus, with an estimate of $dr/dt$ and three dimensional positions from the FP, we can find the density parameter and calculate the associated mass.

We shift the velocities and positions to be relative to A2065, setting it as the centre of the supercluster, since it is the most massive cluster in the region. We then find the smallest radial velocity component consistent with the line-of-sight velocity, giving us an estimate of the mass needed to generate that motion. Essentially, we make the assumption that the cluster's peculiar motion is completely described by the line-of-sight velocity, then take the dot product with the radial unit vector in the supercluster frame to find the associated radial velocity. We find the velocity parameter using equation \eqref{VelParam}, allowing us to then look up the associated value of $\Omega_{\mathrm{s}}$ from our numerical integration results, giving an estimate of the mass enclosed by the shell at the radius of the cluster. By using the smallest possible velocity, we should find a lower bound to the mass, however, the idealizations of the SCM tend to lead to an upper bound \citep{Reisenegger00}. Thus, these results (shown in Table \ref{SCM+FPTab}) should be viewed as optimistic lower bounds.

There are several things worth noting about these results. First, the radial velocities suggest that all the clusters, except A2092 and CL1529+29, are currently collapsing towards A2065. Second, as the distance from A2065 increases, so does the enclosed mass, which is physically what we would expect. The one peculiarity comes in when we look at A2079. As we go to larger radii, the enclosed density is decreasing until we get to A2079, where it suddenly doubles. It also seems rather suspect for A2079 to be so far from A2065 and have such a large component of velocity towards A2065. Given the variability seen in the FP analysis of A2079 (section \ref{sec:dynamics:systematics}), and the peculiarities observed here, we adopt the estimate from A2092 of \mbox{$1.95 \times 10^{16} \, h^{-1} \mathrm{M}_{\sun}$}.

Looking at Table \ref{SCM+FPTab}, we see that the turn-around radius should lie somewhere between A2067 and A2092. In a recent paper, \cite{Karachentsev2014} showed that the mass of a structure can be estimated if the turn-around radius is known via
\begin{equation}
\dfrac{M_{\mathrm{enc}}}{\mathrm{M}_{\sun}} = 2.12 \times 10^{12} \left(\dfrac{R_{0}}{1 \, \mathrm{Mpc}}\right)^{3},
\end{equation}
where $R_{0}$ is the turn-around radius. Using this equation, we can substitute the mass enclosed by A2067 from Table \ref{SCM+FPTab} to see where the turn-around radius should be, and unsurprisingly we find that \mbox{$R_{0} = 12.8 \, h^{-1} \, \mathrm{Mpc}$}.

\cite{D06} find that the density parameter of the critically bound shell is \mbox{$\Omega_{\mathrm{s}} = 2.36$}, meaning any shell with a higher density parameter should be bound in a \mbox{$\Lambda\mathrm{CDM}$} universe. Looking at Table \ref{SCM+FPTab}, we can see that all of the clusters have an associated density parameter higher than the critcial value, which would imply that they should all be part of the bound structure. Due to the idealizations of the SCM, the results of \cite{D06} tend to be optimistic as to the extent of bound structure. The authors note particularly that the presence of ``external attractors'' seem to greatly affect the reliability of their model. Noting that the density parameters for A2092 and CL1529+29 are near the critcal value, predictions of their being part of the bound structure should be viewed with caution, particularly in the case of CL1529+29, which may be influenced by the background supercluster. With these considerations, the SCM+FP method indicates that A2056, A2061, A2065, A2067, and A2089 are part of an extended bound structure, with the potential for A2092 to also be part of that structure.

To test the robustness of this method, we ran 100,000 Gaussian random realizations of the FP results through this procedure in a Monte Carlo (MC) fashion. For each cluster, a Gaussian distribution of FP redshifts was generated using the associated errors as $1\sigma$. A value was then randomly selected from this distribution and the mass estimation procedure repeated. In Fig. \ref{MCHist}, we show histograms for the masses that resulted from the MC procedure. We notice that all these distributions have high mass tails, with a large number of realization yielding similar or smaller masses than the central value which is indicated with the dashed vertical line. If we naively take a standard deviation of the distribution, we get an average of \mbox{$\sim 136$ per cent}. However, given that the value listed for mass in Table \ref{SCM+FPTab} typically lies in the low mass portion of this distribution, we take these mass estimates to be lower bounds. 

Given enough high quality redshift data for the clusters, it would not be unreasonable to get the errors in the distance estimates from the FP down to the 1 per cent level. To examine the effect that would have on the analysis here, we re-ran the MC procedure with the error in $z_{FP}$ set to 1 per cent for each cluster. This reduced the standard deviation in the mass distributions to \mbox{$\sim 40$ per cent} on average, with some cases being as low as \mbox{$\sim 20$ per cent}, showing that these results could be substantially improved with more data. 

\subsection{Application of the caustics method}
\begin{figure}
\centering
\includegraphics[width=1.0\linewidth]{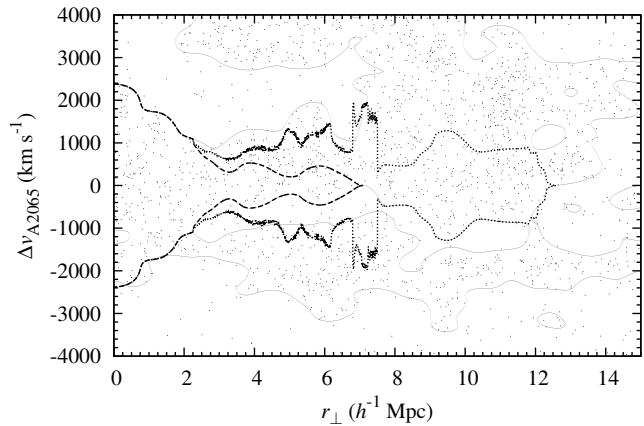}
\caption{Velocity difference versus projected radius. The small dots represent individual galaxies. The solid line is the contour for a density cut off of \mbox{$2.2 \times 10^{-6} \, h \, \mathrm{Mpc}^{-1} \, \mathrm{km}^{-1} \, \mathrm{s}$}. The dotted line is the resulting caustic from method 1, and the dashed line is the resulting caustic from method 2.}
\label{VvsRPerp}
\end{figure}
\cite{Reisenegger00}, building on the work of \cite{Diaferio99}, examined the caustics in the velocity, $v$, versus projected separation, $r_{\perp}$, in order to estimate the mass and bound extent of the SSC. By looking for the density contour in a plot of the aforementioned quantities which forms the ``characteristic trumpet shape'', \cite{Reisenegger00} find the caustics amplitude in velocity space, giving them an estimate of the velocities of galaxies at different distances from the chosen centre. They use this to inform a parametrized form of the spherical collapse model, simultaneously obtaining a mass estimate and a dynamical description of the structure. We choose to apply this method to the CSC in order to obtain a mass estimate based on observational data that is completely independent from the results of the dynamical analysis of Section \ref{sec:dynamics}, giving us a point of comparison and a check of the validity of those results.

Applying the caustics method can be rather subjective and depends on how cleanly the caustics can be discerned from the data. For the SSC, \cite{Reisenegger00} were fairly lucky in that the upper caustic was very well defined in the data, allowing them to simply adopt that for their analysis. Fig. \ref{VvsRPerp} shows that the caustics for the CSC are only really well defined out to \mbox{$r_{\perp} \approx 7 \, h^{-1} \, \mathrm{Mpc}$}, which makes the application of this method a little more difficult.

\begin{figure}
\includegraphics[width=1.0\linewidth]{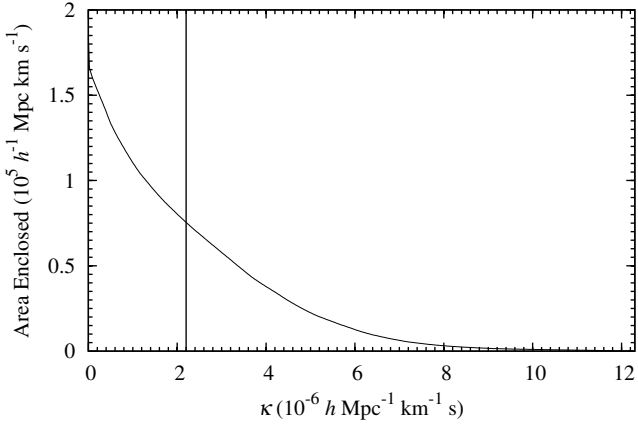}
\caption{The area enclosed by the contour of a particular cut off density, $\kappa$. The vertical line represents the adopted cut off density used in the determination of the caustics, \mbox{$2.2 \times 10^{-6} \, h \, \mathrm{Mpc}^{-1} \, \mathrm{km}^{-1} \, \mathrm{s}$}. In the region around this cut off value, \mbox{$dA/d\kappa$} is approximately constant.}
\label{AvsK}
\end{figure}

Using the same data set as for the virial analysis, we shift everything relative to A2065, our assumed supercluster centre. Taking the spectroscopic redshift of A2065 as a distance indicator, the angular separation on the sky is converted into a projected physical separation. We then calculate the density of the data in Fig. \ref{VvsRPerp} following the methods of \cite{Reisenegger00}, 
\begin{equation}
\label{caustics:den}
f(r_{\perp},v) = \dfrac{1}{N} \sum_{i=1}^{N} \dfrac{1}{h_{r}^{i}h_{v}^{i}} K\left(\dfrac{r_{\perp}-r_{\perp}^{i}}{h_{r}^{i}}, \dfrac{v-v^{i}}{h_{v}^{i}}\right),
\end{equation}
where
\begin{equation}
K(\bmath{t}) = \left\{\begin{array}{lr}
4\mathrm{\pi}^{-1}(1-\left|\bmath{t}\right|^{2})^{3} & \left|\bmath{t}\right| < 1, \\
0 & \left|\bmath{t}\right| > 1, \\
\end{array} \right.
\end{equation}
In equation \eqref{caustics:den}, $h_{r}^{i}$ and $h_{v}^{i}$ are the smoothing lengths along the $r_{\perp}$ axis and the $\Delta v$ axis for the $i$th galaxy, respectively. We adopt the same fixed smoothing lengths as \cite{Reisenegger00}, \mbox{$h_{r} = 1 \, h^{-1} \, \mathrm{Mpc}$} and \mbox{$h_{v} = 500 \, \mathrm{km} \; \mathrm{s}^{-1}$}, and also follow their procedure for calculating the cut off density, arriving at a value \mbox{$\kappa = 1.5 \times 10^{-6} \, h \, \mathrm{Mpc}^{-1} \, \mathrm{km}^{-1} \, \mathrm{s}$}. The density contour at this level encloses over 50 per cent of the area of Fig. \ref{VvsRPerp}. Following the suggested procedure of \cite{Diaferio99}, we adjust our cut off density until it corresponds to the visual impression of the data. Also, following \cite{Reisenegger00}, we look at the area enclosed versus the cut off density, making sure that the value we pick is outside of the range where the area enclosed is rapidly decreasing, and in a region where \mbox{$dA/d\kappa$} is roughly constant. They interpret this as having identified the structure and going to higher density simply selects denser parts of that structure. Fig. \ref{VvsRPerp}, shows that the selected cut off density, \mbox{$\kappa = 2.2 \times 10^{-6} \, h \, \mathrm{Mpc}^{-1} \, \mathrm{km}^{-1} \, \mathrm{s}$}, follows the visual impression of the data, while Fig. \ref{AvsK} shows that it also seems to lie in a region where \mbox{$dA/d\kappa$} is roughly constant.

\begin{figure}
\includegraphics[width=1.0\linewidth]{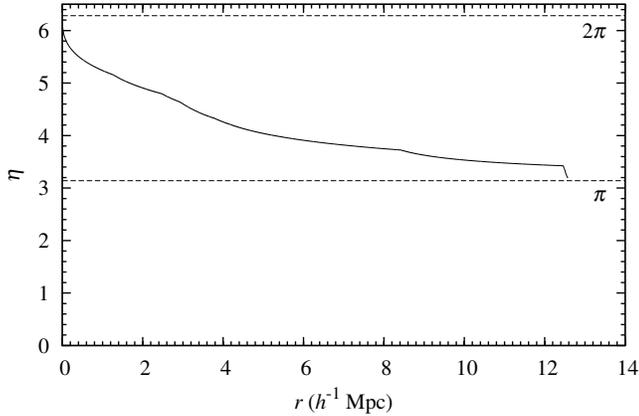}
\caption{SCM parameter $\eta$ vs. $r$ for the CSC as implied by the method 1 caustics. Since a value of $\mathrm{\pi}$ represents shells at turn-around, and a value of \mbox{$2\mathrm{\pi}$} represents a shell that has completely collapsed, we can see that the turn-around radius for the CSC should currently be \mbox{$12.6 \, h^{-1} \, \mathrm{Mpc}$}, and any shell at a smaller radius should be currently collapsing.}
\label{etavsr}
\end{figure}

Finding the caustics amplitude, \mbox{$\mathscr{A}(r_{\perp})$}, is not straight forward. There is clearly some asymmetry in the data (see Fig. \ref{VvsRPerp}) which causes the upper portion of the density contour between \mbox{$3 \, h^{-1} \, \mathrm{Mpc} < r_{\perp} < 7.5 \, h^{-1} \, \mathrm{Mpc}$} to be at higher values of velocity than the lower portion. The asymmetry is most likely rooted in the fact that these structures are not expected to be symmetric themselves. As pointed out by \cite{Reisenegger00}, this asymmetry is not a concern when it comes to the mass estimate since the gravitational field is likely to be more spherically symmetric than the underlying galaxy distribution. The only concern is then choosing which caustics amplitude to follow.

Following the lower boundary, the contour goes to zero around \mbox{$7 \, h^{-1} \, \mathrm{Mpc}$}, which suggests that the caustics should stop there. Following the upper boundary, we find a value of \mbox{$\sim 1000 \, \mathrm{km} \; \mathrm{s}^{-1}$} at \mbox{$7 \, h^{-1} \, \mathrm{Mpc}$}. These would lead to very different results for the mass estimation, motivating us to apply two different methods. For method 1, we follow the upper boundary out to \mbox{$r_{\perp} = 2.25 \, h^{-1} \, \mathrm{Mpc}$}, then follow the lower boundary out to \mbox{$r_{\perp} = 7.5 \, h^{-1} \, \mathrm{Mpc}$}. For method 2, we again follow the upper boundary out to \mbox{$r_{\perp} = 2.25 \, h^{-1} \, \mathrm{Mpc}$}, from \mbox{$2.25 \, h^{-1} \, \mathrm{Mpc} < r_{\perp} \leq 7.5 \, h^{-1} \, \mathrm{Mpc}$} we take the average of the upper and lower boundaries, then from  \mbox{$7.5 \, h^{-1} \, \mathrm{Mpc} < r_{\perp} \leq 12.5$} we follow the lower boundary.  

With the caustics amplitude defined by these methods, we estimate the infall velocity as \citep{Reisenegger00}
\begin{equation}
u(r) \leq u_{b}(r) \equiv \min_{r_{\perp}<r} \dfrac{\mathscr{A}(r_{\perp})}{\left[1-(r_{\perp}/r)^{2}\right]^{1/2}},
\end{equation}
where $u$ is the infall velocity and $r$ is the real space radius. \cite{Peebles93} gives the parametrized solutions for the dynamics of a given shell in the SCM in terms of the phase, $\eta$, which represents the stage of evolution for a shell. A value of \mbox{$\eta = 0$} corresponds to the initial expansion of the shell, \mbox{$\eta = \mathrm{\pi}$} corresponds to the maximum expansion (e.g. turn-around), and a value of \mbox{$\eta = 2\mathrm{\pi}$} corresponds to collapse. With $u(r)$, we calculate the phase for the shell at each radius via
\begin{equation}
-H_{0}t_{1} \dfrac{u}{H_{0}r} = \dfrac{\sin \eta (\eta - \sin \eta)}{(1-\cos \eta)^{2}}.
\end{equation}
Here, $H_{0}t_{1}$ is a dimensionless constant which we can calculate from \citep{Peebles93}
\begin{equation}
H_{0}t(z) = \int_{z}^{\infty} \dfrac{dz}{(1+z)\left[\Omega_{m,0}(1+z)^{3} + \Omega_{\Lambda,0}\right]^{1/2}}.
\end{equation}
Using the redshift of A2065, this gives a value of $H_{0}t_{1} = 0.895$ for the standard \mbox{$\Lambda\mathrm{CDM}$}. With $\eta$ as a function of radius, we calculate the mass as
\begin{equation}
H_{0}M(r) = \dfrac{(H_{0}r)^{3}}{G(H_{0}t_{1})^{2}} \dfrac{(\eta-\sin \eta)^{2}}{(1-\cos \eta)^{3}},
\end{equation}
where $M(r)$ is the mass enclosed by a shell of radius $r$. 

From method 1, the caustics only extend out to \mbox{$\sim 7 \, h^{-1} \, \mathrm{Mpc}$}, preventing us from being able to find the mass enclosed beyond that radius. At \mbox{$r = 6.98 \, h^{-1} \, \mathrm{Mpc}$} we find a mass of \mbox{$2.05 \times 10^{15} \, h^{-1} \, \mathrm{M}_{\sun}$}. This is consistent within errors with the sum of the masses of A2065 and A2056, which are the only clusters within that radius.

From method 2, we get an estimate of \mbox{$1.02 \times 10^{16} \, h^{-1} \, \mathrm{M}_{\sun}$} within a radius of \mbox{$r = 12.46 \, h^{-1} \, \mathrm{Mpc}$}. Fig. \ref{etavsr} shows that the current turn-around radius should be \mbox{$r_{\mathrm{TA}} \approx 12.6 \, h^{-1} \, \mathrm{Mpc}$}. This means that A2056, A2061, A2065, A2067, and A2089 are within the turn-around radius and should be part of an extended bound structure. Considering the SCM+FP results (see Table \ref{SCM+FPTab}), we see that A2067 is at a distance of \mbox{$12.5 \, h^{-1} \, \mathrm{Mpc}$} and is essentially at turn-around. The mass predicted to be within that radius by the SCM+FP is \mbox{$9.1 \times 10^{15} \, h^{-1} \, \mathrm{M}_{\sun}$}, which is remarkably consistent with the results of the caustics analysis. 

\section{N-body Simulations}
\label{sims}
We have performed simulations of the CSC using the same code as P13 in order to assess the likelihood that the CSC is a gravitationally bound supercluster. In total 126,000 simulations were performed, exploring a wide variety of possible initial conditions; reducing the initial expansion rate of the clusters below the Hubble flow based on the enclosed mass using the SCM, assigning only line-of-sight velocities predicted by the fundamental plane, and including different amounts of inter-cluster mass based on the range of values presented in section \ref{scmasssec}. Performing large numbers of simulations allows us to assess the likelihood that the structure is bound, by taking into account the uncertainties in positions, masses and velocities. Below we choose to present the results of only some of these simulations, since many have similar initial conditions with similar results.

\subsection{Simulations with no inter-cluster mass}
\label{sim:noICM}
\begin{table*}
\centering
\caption{Simulation results for Set \#1 and \#2. Column (1) lists the cluster pair being examined. Column (2) shows the results for Set \#1 based on a comparison of the kinetic and potential energies at the end of the simulation. Column (3) lists the number of simulations in Set \#1 which the pair has a close encounter (coming within 3 Mpc of each other), and column (4) lists our estimate for the likelihood of the pair being gravitationally bound. Columns (5), (6) and (7) are the same as columns (2), (3), and (4) respectively, only for Set \#2. If no data are listed for a cluster pair in a particular simulation set, that pair is not of interest having no chance of being gravitationally bound.}
\label{TabSet1-2}
\begin{threeparttable}
\begin{tabular}{lcccccc}
\hline
${}$ & \multicolumn{3}{c}{Set \#1} & \multicolumn{3}{c}{Set \#2} \\
Cluster Pair & Results & Close encounters & Per cent likelihood & Results & Close encounters & Per cent likelihood \\
\hline
A2061/A2067 & $28 \pm 3.3$ & $26 \pm 2.6$ & $11.2 \pm 1.3$ & $134 \pm 10.5$ & $151 \pm 9.7$ & $60.4 \pm 3.4$\tnote{a} \\
A2061/A2089 & -- & -- & -- & $8 \pm 4$ & $0 \pm 0$ & $3.2 \pm 1.6$\\
A2065/A2067 & -- & -- & -- & $45 \pm 25$ & $2 \pm 2$ & $18 \pm 10$\\
A2065/A2079 & $12 \pm 3.2$ & $10 \pm 0.8$ & $4.8 \pm 1.3$ & -- & -- & -- \\
A2065/A2089 & $8 \pm 1.9$ & $4 \pm 0.5$ & $3.2 \pm 0.8$ & $164 \pm 6.2$ & $105 \pm 7.1$ & $65.6 \pm 2.5$ \\
A2067/A2089 & -- & -- & -- & $20 \pm 10$ & $2 \pm 1$ & $8.0 \pm 3.9$\\
A2079/A2089 & $4 \pm 0.0$ & $5 \pm 0.2$ & $1.6 \pm 0.0$ & -- & -- & --\\
A2079/A2092 & -- & -- & -- & $69 \pm 11.6$ & $4 \pm 9.9$ & $27.6 \pm 4.6$ \\

\hline
\end{tabular}
\begin{tablenotes}
\item[a] This likelihood is calculated based on the number of close encounters. Tidal friction between clusters would likely lead to mergers after close encounters.
\end{tablenotes}
\end{threeparttable}
\end{table*}

The first simulations performed (Set \#1) reproduce those of by P13. The only difference was the inclusion of the clusters A2079, and A2092 to be completely consistent with the cluster sample of B13. The results of these simulations are presented in columns (2), (3) and (4) of Table \ref{TabSet1-2}, where the numbers are out of a total of 250, with the errors being the $1\sigma$ variations seen in the 20 Gaussian random mass realizations (see P13 for details). These are entirely consistent with the results of P13, showing very little chance of any extended bound structure. This is not surprising given that the additional clusters are both fairly separated from the four included by P13 (A2061, A2065, A2067, and A2089) and don't add a substantial amount of mass. 

The most interesting results for simulations with no inter-cluster mass were the ones informed entirely by the FP results (Set \#2). For these simulations the positions of the clusters were determined using a Gaussian random realization of the FP redshift where the errors in Table \ref{tab:FP} were used as $1\sigma$. We then used equation \eqref{eq:V_pec} with the varied $z_{FP}$'s to assign only a line-of-sight velocity to each cluster. In this manner 250 sets of initial conditions were generated. Each realization was run using the central values of mass presented in Table \ref{CMSep}, and for 20 random Gaussian realizations of masses for the clusters. The results of the simulations are presented in columns (5), (6) and (7) of Table \ref{TabSet1-2}. We now see that A2061/A2067 and A2065/A2089 have a greater than 50 per cent chance of being gravitationally bound pairs, and A2067 also has some chance of being bound to A2065 meaning it may act as a ``bridge'' connecting the pairs. 

The fact that simulations show a significant probability of these clusters interacting, strongly suggests that the inclusion of some inter-cluster matter should lead to extended bound structure. As we have shown in Section \ref{scmasssec}, there is likely a significant inter-cluster matter component. 

Since both of these simulation sets were intended to be consistent with B13, neither of these included A2056 or CL1529+29. Given the positions of these clusters and masses (see Fig. \ref{fig:clusterpos}, Table \ref{CMSep}, and Table \ref{tab:FP}), it is likely that A2056 would show up as bound to A2065, while CL1529+29 would not be bound to any part of the structure. The simulations with inter-cluster mass that include these two clusters would also seem to support these conclusions (see discussion below).

\subsection{Simulations with inter-cluster mass}
\label{sims:ICM}
\begin{table*}
\centering
\caption{Simulation results for Halo \#1. Column (1) lists the cluster name. Column (2) has the number of simulations in which the particular cluster was gravitationally bound to Halo \#1. Column (3) lists the close encounters with the supercluster centre, and column (4) has the per cent likelihood that the cluster is bound.  Columns (5), (6), and (7) repeat columns (2), (3), and (4) respectively for Halo \#2. Note that A2065 is not listed since it is placed at the centre of the supercluster, in the middle of the halo.}
\label{TabHalo1-2}
\begin{threeparttable}
\begin{tabular}{lcccccc}
\hline
${}$ & \multicolumn{3}{c}{Halo \#1} & \multicolumn{3}{c}{Halo \#2} \\
Cluster & Results & Close encounters\tnote{a} & Per cent likelihood & Results & Close encounters\tnote{a} & Per cent likelihood\\
\hline
A2056 & $238 \pm 2$ & $151 \pm 51$ & $95.2 \pm 0.8$ & $165 \pm 5$ & $150 \pm 15$ & $66.0 \pm 1.9$ \\
A2061 & $193 \pm 7$ & $0 \pm 1$ & $77.2 \pm 2.8$ & $79 \pm 19$ & $0 \pm 1$ & $31.6 \pm 7.6$ \\
A2067 & $236 \pm 4$ & $3 \pm 8$ & $94.4 \pm 1.8$ & $152 \pm 15$ & $4 \pm 5$ & $60.8 \pm 5.9$ \\
A2079 & $245 \pm 4$ & $0 \pm 0$ & $98.0 \pm 1.6$ & $61 \pm 14$ & $0 \pm 0$ & $24.4 \pm 5.6$ \\
A2089 & $249 \pm 1$ & $92 \pm 7$ & $99.6 \pm 0.2$ & $223 \pm 2$ & $93 \pm 9$ & $89.2 \pm 0.9$ \\
A2092 & $170 \pm 8$ & $0 \pm 0$ & $68.0 \pm 3.1$ & $29 \pm 8$ & $0 \pm 0$ & $11.6 \pm 3.3$ \\
CL1529+29 & $2 \pm 1$ & $0 \pm 0$ & $0.8 \pm 0.5$ & $0 \pm 0$ & $0 \pm 0$ & $0.0 \pm 0$ \\
\hline
\end{tabular}
\begin{tablenotes}
\item[a] Close encounters with A2065, the assumed centre of the CSC.
\end{tablenotes}
\end{threeparttable}
\end{table*}

The addition of inter-cluster mass to the simulations requires some modification to the code. Large scale N-body simulations, such as the Millennium simulation \citep{Springel05b}, show that dark matter forms a filamentary structure with clusters forming at the intersections of filaments. However, there is only contradictory observational evidence for these dark matter filaments \citep{Gray02,Heymans08,Dietrich12} and no studies looking for them have been performed on the CSC. In addition, including filaments in the simulations would greatly increase computation time, reducing the number of realizations we could explore. This motivated us to take a simpler approach which should still be an effective means of exploring the impact of inter-cluster matter.

For these simulations we chose A2065 as the centre, and added in inter-cluster mass as a virtual halo that follows a density profile of the form presented in \cite{Navarro97} (the NFW profile). The acceleration of each cluster was then updated as the sum of the accelerations from all other clusters and the enclosed mass from the virtual halo. In this manner, we explored the effects of inter-cluster mass by only including one additional calculation per cluster, per calculation step. This allowed us to perform large numbers of simulations exploring a wide variety of initial conditions, since the addition of the halo negligibly increased the time of a simulation run.

Of course, a simplified model has its draw backs. The spherically symmetric halo will only enhance accelerations towards the chosen centre of the supercluster leading to different trajectories than if filaments were included. Since there are no actual particles associated with the halo, there are no tidal friction effects which could lead to a larger extended bound region. For this reason, we conclude that the simulations results are still somewhat of a lower limit to the extent of bound structure.

We have performed simulations using five different possible values for the inter-cluster mass halo, with a variety of choices for the initial conditions. Overall, as expected, the more inter-cluster mass included, the higher the likelihood of seeing extended bound structure. For this reason, we only present the results of simulations that include an inter-cluster mass halo of \mbox{$3.8 \times 10^{16} \, h^{-1} \, \mathrm{M}_{\sun}$} within \mbox{$\sim 11 \, h^{-1} \, \mathrm{Mpc}$} of A2065 (Halo \#1), and \mbox{$1.0 \times 10^{16} \, h^{-1} \, \mathrm{M}_{\sun}$} within \mbox{$\sim 12.5 \, h^{-1} \, \mathrm{Mpc}$} of A2065 (Halo \#2). Halo \#1 was motivated by the work of \cite{small3}, and the fact that the mass lies roughly in the middle of the range established in section \ref{scmasssec}, while Halo \#2 was motivated by both the caustics and SCM+FP analyses. The results presented are based on the kinetic and potential energies at the end of the simulation, where the potential energy is calculated based on the enclosed mass (clusters + halo).

Columns (2), (3) and (4) of Table \ref{TabHalo1-2} show the results for Halo \#1. With this particular halo, there is a \mbox{$\sim 95$ per cent} chance of A2056, A2065, A2067, A2079, and A2089 being gravitationally bound, and a \mbox{$\sim 68$ per cent} chance of all clusters other than CL1529+29 being bound. This would seem to lend credence to the idea that applying the virial theorem on the supercluster scale may simply be finding the mass required to bind the structure.

Columns (5), (6), and (7) of Table \ref{TabHalo1-2} show the results for Halo \#2. Even by cutting the amount of inter-cluster mass by a factor of \mbox{$\sim 4$}, we still find a \mbox{$\sim 61$ per cent} chance of A2056, A2065, A2067, and A2089 being a gravitationally bound structure. The substantial reductions in likelihood can be explained by considering the FP redshifts ($z_{FP}$) listed in Table \ref{tab:FP}. A2061 has an FP redshift of 0.0727, while A2065 has an FP redshift of 0.0730, placing them at virtually the same distance. Due to this the relative motion will be almost entirely tangential, and is estimated to be fairly large \mbox{($\sim 1000 \, \mathrm{km} \, \mathrm{s}^{-1}$)}. Without a substantial inter-cluster matter component, it is easy for this cluster to escape. We expect to see a similar, though less substantial, effect for A2056, A2067, and A2089, yet another reason why the simulation results should be regarded as a lower limit to the extent of bound structure in the region.

These simulations demonstrate that given a reasonable amount of inter-cluster matter, there is a significant chance of finding extended bound structure in the CSC. \cite{Proust06} find that in the Shapley supercluster, the inter-cluster galaxies may contribute twice as much mass as the cluster galaxies. The sum of the masses of the clusters in the CSC (excluding CL1529+29) is \mbox{$5.9 \times 10^{15} \, h^{-1} \, \mathrm{M}_{\sun}$}, and if the CSC is similar to Shapley, it would not be unreasonable to expect at least \mbox{$1.0 \times 10^{16} \, h^{-1} \, \mathrm{M}_{\sun}$} in the inter-cluster region.

\section{Discussion}

It is fairly straightforward to place firm limits on the possible mass of the CSC. Application of the virial theorem sets an upper bound of \mbox{$1.2 \times 10^{17} \, h^{-1} \, \mathrm{M}_{\sun}$}, and summing the cluster masses (excluding CL1529+29) gives a lower bound of \mbox{$5.9 \times 10^{15} \, h^{-1} \, \mathrm{M}_{\sun}$}. These bounds cover a significant range, and it is unlikely that either method provides an accurate assessment of the mass. Given the probability that inter-cluster mass contributes significantly to the total mass of the CSC \citep{Proust06}, it is all but certain that the mass is larger than that obtained by summing the cluster masses. Similarly, the virial theorem assumes a bound virialised structure extending over a significant range in velocity space, and therefore very likely overestimates the total mass of the CSC. The dynamical analysis and resulting simulations and mass estimation, as well as the independent method of mass estimation provided by the caustics method, are intended to provide a more physically meaningful assessment of its mass and bound extent.

The FP analysis indicates significant extended bound structure in the CSC, suggesting two bound regions that have reached turnaround and are in collapse: the first consisting of A2065, 2056 and 2089, and the second consisting of 2061 and 2067, and possibly also 2092. There is also some indication that these regions may not be dynamically isolated, and the CSC may consist of a single collapsing core of at least five clusters. Simulations informed by the FP results, and including no inter-cluster mass, support these conclusions, indicating a significant likelihood \mbox{($> 50$ per cent)} of each of the two cores being gravitationally bound, and some likelihood \mbox{($\sim 18$ per cent)} of the two regions being bound to each other. The errors on the FP distance estimates are quite low and appear to be free of any systematic bias, and the resulting cluster peculiar velocities are consistent with what might be expected for a bound structure of this type (see discussion in B13). However it is important to consider that the distance errors are still too large compared to the peculiar velocities to give a definitive analysis of the dynamics, so the results from extensive simulations that account for these errors are key in assessing the validity of the FP results and the conclusions we draw from them.

The two most physically meaningful methods of mass estimation are the SCM+FP, which is based on the FP results but also accounts for the errors in the distance determinations, and the caustics method, which provides an independent method of mass estimation based on observational data. The SCM+FP analysis indicates a bound region with a mass of \mbox{$0.91 \times 10^{16} \, h^{-1} \, \mathrm{M}_{\sun}$} within a turn-around radius of \mbox{$12.5 \, h^{-1} \, \mathrm{Mpc}$}, while the caustics method indicates a bound region with a mass of \mbox{$1.02 \times 10^{16} \, h^{-1} \, \mathrm{M}_{\sun}$} within a turn-around radius of \mbox{$12.6 \, h^{-1} \, \mathrm{Mpc}$}.  The remarkable agreement between the mass estimates and turn-around radii from these methods is particularly significant. Given the fundamental differences between the data sources and methods of analysis, it is highly unlikely that this consistency is the result of some underlying systematic bias. Consequently, each of these methods can be viewed as independently verifying the other. Since the SCM+FP method is informed by the FP, this consistency suggests that our method of calibration is valid, and that the FP analysis accurately assesses the internal dynamics of the CSC. Taken together, these results indicate that the CSC contains a single bound core containing: A2056, A2061, A2067, A2089, and A2065, which has reached turnaround and is in collapse with a mass of \mbox{$\sim 1 \times 10^{16} \, h^{-1} \, \mathrm{M}_{\sun}$}.

The results of Section \ref{sims:ICM} demonstrate that reasonable assumptions about inter-cluster mass result in a significant likelihood of extended bound structure in the CSC. The low mass halo simulations assume an inter-cluster matter component that requires no dark matter, and results in a \mbox{$\sim 60$ per cent} chance of A2056, A2065, A2067 and A2089 being bound, with a significantly lower likelihood that A2061 is part of the structure. Given the large tangential peculiar velocity of A2061 that results from the FP analysis, and its proximity to A2067, the simulations probably under-estimate the likelihood that they are a bound pair. The higher mass halo simulations would not be unreasonable assuming the presence of inter-cluster dark matter, and in that case the likelihood of the aforementioned clusters forming a bound structure is \mbox{$\sim 95$ per cent}, with A2061 showing a lower probability for the reasons given above. We also note that, if there is this much inter-cluster mass present, there is a good chance \mbox{($\sim 68$ per cent)} that A2092 is also part of the structure.

Combining the results of the FP analysis, caustics method, SCM+FP, and simulations, we conclude that A2056, A2061, A2065, A2067, and A2089 comprise a gravitationally bound supercluster core. Should there exist an inter-cluster dark matter component, A2092 may also be part of this structure. This work provides the most conclusive evidence to date that the CSC is a bound supercluster similar to the SSC, and suggests that such structures may be found elsewhere in the Universe. 

\section*{Acknowledgements}
This research has made use of the NASA/IPAC Extragalactic Database (NED) which is operated by the Jet Propulsion Laboratory, California Institute of Technology, under contract with the National Aeronautics and Space Administration (NASA). Portions of this research were funded by grants from NASA through the Maine Space Grant Consortium (MSGC). This research has also made use of the Sloan Digital Sky Survey (SDSS) data. Funding for the SDSS and SDSS-II has been provided by the Alfred P. Sloan Foundation, the Participating Institutions, the National Science Foundation, the U.S. Department of Energy, the National Aeronautics and Space Administration, the Japanese Monbukagakusho, the Max Planck Society, and the Higher Education Funding Council for England. The SDSS Web Site is http://www.sdss.org/. The SDSS is managed by the Astrophysical Research Consortium for the Participating Institutions. The Participating Institutions are the American Museum of Natural History, Astrophysical Institute Potsdam, University of Basel, University of Cambridge, Case Western Reserve University, University of Chicago, Drexel University, Fermilab, the Institute for Advanced Study, the Japan Participation Group, Johns Hopkins University, the Joint Institute for Nuclear Astrophysics, the Kavli Institute for Particle Astrophysics and Cosmology, the Korean Scientist Group, the Chinese Academy of Sciences (LAMOST), Los Alamos National Laboratory, the Max-Planck-Institute for Astronomy (MPIA), the Max-Planck-Institute for Astrophysics (MPA), New Mexico State University, Ohio State University, University of Pittsburgh, University of Portsmouth, Princeton University, the United States Naval Observatory, and the University of Washington.

\bibliographystyle{mn2e}
\bibliography{PearsonBatiste-TheCSC}

\label{lastpage}

\end{document}